\documentclass[final,5p,times]{article}

\usepackage{multirow,booktabs}

\usepackage{amsmath}

\usepackage{listings}
\usepackage{float}
\usepackage{rccol}
\usepackage[table]{xcolor}
\usepackage{algorithm2e}
\usepackage{wrapfig}
\usepackage{graphicx}
\usepackage{plain}
\usepackage{authblk}
\usepackage{amsmath}

\usepackage[normalem]{ulem}

\usepackage{xr}
\usepackage{setspace}
\usepackage[margin=0.7in]{geometry}
\usepackage{amssymb}
\usepackage{makecell}

\begin{document}
\noindent \large{\textbf{Multiscale and multimodal network dynamics underpinning working memory}}
\\
\\
Andrew C. Murphy\textsuperscript{1,2}, Maxwell A. Bertolero\textsuperscript{1}, Lia Papadopoulos\textsuperscript{3}, David M. Lydon-Staley\textsuperscript{1}, and Danielle S. Bassett\textsuperscript{1,3-6,*}
\\
\\
\textsuperscript{1}Department of Bioengineering, School of Engineering \& Applied Sciences, University of Pennsylvania, Philadelphia, PA 19104, USA
\\
\textsuperscript{2}Perelman School of Medicine, University of Pennsylvania, Philadelphia, PA 19104, USA
\\
\textsuperscript{3}Department of Physics \& Astronomy, School of Arts \& Sciences, University of Pennsylvania, Philadelphia, PA 19104, USA
\\
\textsuperscript{4}Department of Neurology, Perelman School of Medicine, University of Pennsylvania, Philadelphia, PA 19104, USA
\\
\textsuperscript{5}Department of Psychiatry, Perelman School of Medicine, University of Pennsylvania, Philadelphia, PA 19104, USA
\\
\textsuperscript{6}Department of Electrical \& Systems Engineering, School of Engineering \& Applied Sciences, University of Pennsylvania, Philadelphia, PA 19104, USA
\\
\textsuperscript{*}To whom correspondence should be addressed: dsb@seas.upenn.edu

\clearpage
\section*{Abstract}
As a key component of executive function, working memory allows information to be stored and manipulated over short time scales. Performance on laboratory and ecological working memory tasks is thought to be supported by the frontoparietal system, the default mode system, and interactions between them. Yet, little is known about how these systems and their interactions might relate to, or perhaps even explain, individual differences in working memory performance. Here, we address this gap in knowledge using functional magnetic resonance imaging data acquired during the performance of a 2-back working memory task, as well as diffusion tensor imaging data collected in the same individuals as part of the Human Connectome Project. We show that the strength of functional interactions between the frontoparietal and default mode systems during task engagement is inversely correlated with working memory performance. Further, we demonstrate that the strength of functional interactions between these two systems is modulated by the activation of frontoparietal regions but not default mode regions. To gain a deeper understanding of these processes, we used an unsupervised clustering algorithm informed by an explicit model of network architecture to identify two distinct subnetworks of the frontoparietal system, and we subsequently demonstrate that these subnetworks also display distinguishable patterns of gene expression. We found that activity in one subnetwork was positively associated with the strength of the functional interaction between the frontoparietal and default mode systems, while activity in the second subnetwork was negatively associated. Using the diffusion imaging data, we then demonstrate that the pattern of structural linkages between these subnetworks explains their differential capacity to influence the strength of the functional interaction between the frontoparietal and default mode systems. To determine whether these correlative observations could provide a mechanistic account of large-scale neural underpinnings of working memory, we built a computational model of the system composed of simplified coupled oscillators. We demonstrate that modulating the relative amplitude of the subnetworks in the model causes the expected change in the strength of the functional interaction between the frontoparietal and default mode systems, thereby offering support for a candidate mechanism in which subnetwork activity tunes functional connectivity. Broadly, our study presents a holistic account of how regional activity, functional connections, and structural linkages together support individual differences in working memory in humans. 

\clearpage
\newpage
\section{Introduction}

Working memory supports the short-term storage of information, thereby facilitating its further manipulation and processing \cite{Baddeley2003}. Individual differences in working memory performance have been associated with differences in the recruitment of distinct cognitive systems, and in the functional interactions between such systems \cite{Keller2015,Liu2018,Bertolero2018}. Heavily implicated in working memory function is a group of regions in frontoparietal cortex involved in cognitive control, among other functions \cite{Pessoa2002,Palva2010}, as well as a group of regions comprising the so-called default mode system, which is most notable for its strong activation in the human resting state. Interestingly, these two systems are thought to have opposite effects on working memory: frontoparietal activity is vital for directing attention to external stimuli \cite{Ptak2012}, while default mode activity is important for internally-directed cognition \cite{Andrews-Hanna2012}. Furthermore, these two systems tend to be in functional \emph{competition} during tasks with high working memory load; in such tasks, the activity of the two systems is anti-correlated \cite{ClareKelly2008}.  

Despite extensive study, the functional role of competition between the frontoparietal and default mode systems remains unclear. Over the past decade, several hypotheses have been put forward. For example, one notable hypothesis suggests that competition between the frontoparietal and default mode systems during certain tasks enables maximally disjunctive levels of activity \cite{ClareKelly2008}. Put more simply, this inter-system competition might allow for a pattern of whole brain dynamics characterized by heightened activity in the frontoparietal system coupled with decreased activity in the default mode system. Explaining such a pattern of dynamics is particularly important in light of evidence that it favors improved working memory performance \cite{Anticevic2010}. A second notable hypothesis suggests that competition may enable a toggling between (i) brain states eliciting default mode activity for introspective processes and (ii) brain states eliciting frontoparietal activity for externally-directed attentional processes \cite{Fornito2012}. A third hypothesis suggests that inter-system competition could prove useful for continuously modulating the brain's response to task complexity; while activity in the frontoparietal system scales with the complexity of the task, activity in the default mode system decreases with the complexity of the task \cite{Bertolero2015}. It is worth noting that -- while of course conceptually interesting -- none of these hypotheses offer particularly explicit mechanisms for competition. 

Its functional role aside, the anatomical substrates and extent of the competition between the frontoparietal and default mode systems remain unclear. Structurally, it is well-known that the two systems are quite distinct in terms of their topological role within the connectome: the default mode system is part of the so-called rich club of the structural connectome, which is a set of densely interconnected high degree nodes, while the frontoparietal system is part of the so-called diverse club, which is a set of densely interconnected nodes with diverse connectivity across all putative cognitive systems \cite{Bertolero2017}. It is intuitively plausible that such distinct placements within the larger whole-brain network could constrain or define the roles that each system can play in inducing certain types of dynamics \cite{Gu2015a,Cornblath2018}. In addition to its anatomical substrates, the anatomical extent of the competition remains unclear. Is the entirety of the default mode system competitive with the entirety of the frontoparietal system? Are these systems in fact homogeneous or could they contain subnetworks with distinct dynamics? Initial evidence suggests that the frontoparietal system can be separated into two subnetworks: one more connected to the default mode system and one more connected to the dorsal attention system \cite{Dixon2018}. Similar evidence suggests that the default mode system can be separated into several subnetworks \cite{Andrews-Hanna2010} supporting distinct processes in the internal generation of thought and the local processing of information \cite{Christoff2016}. A careful investigation into the nature of these system-specific subnetworks may inform more mechanistic models of competition and its relevance for working memory performance. 

Here, we seek to perform a deeper study of individual differences in working memory function and its relation to multimodal neural phenotypes including regional activity, inter-regional connectivity, and structural linkages. We also seek to complement data-driven analysis of empirical measurements with biologically-motivated computational modeling to probe the validity of our explanations and posited mechanisms. Specifically, in this study we use functional magnetic resonance imaging data collected from 644 healthy adult human participants in the Human Connectome Project during the performance of a 2-back working memory task. To address potential structural drivers of our findings, we also use diffusion tensor imaging data acquired in the same participants. We begin with a broad investigation into the relations between behavior, activity within the frontoparietal and default mode systems, and competition between the frontoparietal and default mode systems. We then use a model-based machine learning algorithm to uncover functional groups or \emph{subnetworks} within the entire frontoparietal system, which we subsequently find to display distinguishable patterns of gene coexpression. We show that subnetwork activity tracks the strength of the functional interaction between the frontoparietal and default mode systems, in a manner that is consistent with the underlying pattern of structural linkages between them. Lastly, -- drawing on methods from complex systems physics and non-linear dynamical systems theory -- we build a computational model of the system and its dynamics as a collection of coupled oscillators, with parameters and coupling architecture informed by the biological evidence we uncover. We use the model to probe which features of the biology are able to produce the observed phenomena, and to test our hypotheses about the relationships between activity, anatomical connectivity, and functional interactions.

\section{Materials and Methods}

\subsection{Imaging data acquisition and preprocessing}

\subsubsection{Data acquisition}

For each subject in the Human Connectome Project (HCP) S900 release \cite{VanEssen2013}, we extracted the task-based functional magnetic resonance imaging data acquired during the performance of the n-back working memory task, a resting state functional magnetic resonance imaging scan, a high-resolution anatomical scan, and a diffusion tensor imaging scan. In this release, 644 subjects contained all 4 data types, all four resting-state scans, and BedpostX diffusion data. Participants were 346 females, mean age (std) = 28.6 (3.68) years. No additional exclusion criteria were applied. All analyses were performed in accordance with the relevant ethical regulations of the WU-Minn HCP Consortium Open Access Data Use Terms. Informed consent was obtained in writing from all participants.

The acquisition parameters for each data type are as follows. The parameters for the acquisition of high-resolution structural scan were: TR = 2400 ms, TE = 2.14 ms, TI = 1000 ms, flip angle = 8 deg, FOV = 224 $\times$ 224 mm, voxel size = 0.7 mm isotropic, BW = 210 Hz/Px, acquisition time = 7:40 minutes. Functional magnetic resonance images were collected during both rest and task with the following parameters: TR = 720 ms, TE = 33.1 ms, flip angle = 52 deg, FOV = 208$\times$180 mm, matrix = 104$\times$90, slice thickness = 2.0 mm, number of slices = 72 (2.0 mm isotropic), multi factor band = 8, echo spacing = 0.58 ms. Diffusion tensor images were collected with the following parameters: TR = 5520 ms, TE = 89.5 ms, flip angle = 78 deg, refocusing flip angle = 160 deg, FOV = 210$\times$180, matrix = 168$\times$144, slice thickness = 1.25 mm, number of slices = 111 (1.25 mm isotropic), multiband factor = 3, echo spacing = 0.78 ms, $b$-vaues = 1000, 2000, and 3000 s/mm2. 

\subsubsection{Working memory task and associated measure of behavior}

We focused our analyses on data acquired during the n-back task, due to its reliable recruitment of the executive system \cite{Satterthwaite2013}. The working memory task was presented at two different levels of difficulty: 0-back and 2-back. For both levels, subjects were presented with a stream of images taken from the following four categories: faces, places, tools, or body parts. The latter images presented body parts that were whole (non-mutilated); no images contained nudity. In the 0-back condition, subjects were meant to respond positively during every image presentation. In the 2-back condition, subjects were meant to respond positively if the present image was identical to the image presented two previously. The task was divided into two runs, each run composed of 8 task blocks and 2 fixation blocks \cite{Mason2009}. Fixation blocks lasted for 15 seconds each. Each task block consisted of 10 trials, where a stimulus was presented for 2 seconds, followed by a 500 ms ITI (2.5 seconds total per trial) \cite{Mason2009}. Each block begins with a 2.5 second cue indicating the 0-back or 2-back condition. During the 0-back condition, working memory loads are minimal. Within each run, half of the task blocks used the 2-back paradigm, and half used the 0-back paradigm. Additionally, within a run, each stimulus type (images from a single category) was presented in a different block. To estimate behavioral performance, we calculated the accuracy of responses across all image categories separately for 0-back and 2-back conditions. We chose to focus on accuracy due to its interpretability \cite{Bertolero2018}. However, we also considered d-prime \cite{Satterthwaite2013,Kane2007} and demonstrate that our main results hold when using this metric in place of accuracy (see Supplementary Materials). 

\subsubsection{Processing of functional magnetic resonance imaging data}

For both resting-state and task functional connectivity, CompCor, with five principal components from the ventricles and white matter masks, was used to regress out nuisance signals from the time series. In addition, the 12 detrended motion estimates provided by the Human Connectome Project were regressed out from the time series. The mean global signal was removed and then time series were band-pass filtered from 0.009 to 0.08 Hz. Finally, frames with greater than 0.2 mm frame-wise displacement or a derivative root mean square (DVARS) above 75 were removed as outliers. Sessions composed of greater than 50 percent outlier frames were not further analyzed.

We chose to regress out the global signal from the time series because it has been shown to remove motion signal and global scanner noise. We also note that the mathematics of global signal regression does not necessitate a specific spatial distribution of negative correlations \cite{Murphy2017}, and our claims regard the relative strengths of connectivity between networks rather than their sign. Moreover, the processing pipeline used here has been suggested to be ideal for removing false relations between connectivity and behavior \cite{Siegel2017}. Finally, we note that our main results hold when using wavelet coherence as a measure of functional connectivity (see Supplementary Materials); this measure is bounded between 0 and 1.

\subsubsection{Preprocessing of diffusion tensor imaging data}

For the diffusion imaging, the Human Connectome Project applied intensity normalization across runs, the TOPUP algorithm for EPI distortion correction, the EDDY algorithm for eddy current and motion correction, gradient nonlinearity correction, calculation of gradient $b$-value/$b$-vector deviation, and registration of mean $b$0 to native volume T1w with FLIRT. BBR+bbregister and transformation of diffusion data, gradient deviation, and gradient directions to 1.25 mm structural space were also applied. The brain mask is based on the FreeSurfer segmentation. The BedpostX (Bayesian Estimation of Diffusion Parameters Obtained using Sampling Techniques) output was then calculated, where the `X' stands for modeling crossing fibers. Markov Chain Monte Carlo sampling was used to build probability distributions on diffusion parameters at each voxel. The process creates all of the files necessary for running probabilistic tractography. Using the Freesurfer recon-all data computed by the Human Connectome Project, the fsaverage5 space cortical parcellation was registered to the subject's native cortical white matter surface and then transformed to the subject's native diffusion volume space. From these data, we derived seeds and targets for probabilistic tractography, which we ran with the FSL probtrackx2 algorithm using 1000 streams initiated from each voxel in a given parcel. 

\subsubsection{Whole brain functional parcellation}

We parcellated the brain into 400 discrete and non-overlapping regions of interest using the Schaefer atlas (fslr32k surface) \cite{Schaefer2017}. Notably, the Schaefer atlas was originally developed in the same HCP data that we study here, and it yields a functional demarcation of both the default mode and the frontoparietal systems. Of course other functionally defined atlases exist, but they are less ideal for our purposes for several reasons; the Power atlas \cite{Power2011a} does not provide full cortical coverage, and the Gordon \cite{Gordon2016} and Brainnetome \cite{Fan2016} atlases are lower spatial resolution including 333 and 246 regions, respectively. The Schaefer atlas provides an assignment of each region to one of 17 putative cognitive systems: two visual, two somatomotor, two dorsal attention, two salience/ventral attention, one limbic, three frontoparietal, three default mode, and one temporo-parietal system. To ensure that the granularity of the data was consistent with the granularity of our hypotheses, we collapsed these 17 systems into 8 systems by combining individual systems that belonged to the same cognitive system; that is, we combined the two visual systems into a single system, the two somatomotor systems into a single system, the two dorsal attention systems into a single system, the two salience systems into a single system, the three frontoparietal systems into a single frontoparietal system, and the three default mode systems into a single system. 

\subsection{Analysis of functional magnetic resonance imaging data}

\subsubsection{Estimation of functional activation}

To measure functional activation during task performance, we employed a general linear model (GLM). We collated time series from each brain region during either the 0-back condition or the 2-back condition. For each of the two conditions, we formulated a regressor that indicated the volumes in the time series during which the subject was engaged in the task (not in a fixation block). Treating the 0-back and 2-back time series separately, we convolved this regressor with the canonical hemodynamic response function using the SPM (Wellcome Centre for Human Neuroimaging, London, UK) function \emph{spm\_hrf}, to create a matrix $M_{conv}\in \mathcal{R}^{N\times T}$, where \emph{N} is the number of regions and \emph{T} is the number of time points. We collated observed regional time series to create a matrix $M_{TS} \in \mathcal{R}^{N\times T}$. We then used the MATLAB function \emph{mldivide} to solve the element-wise multiplication equation $M_{TS} = \gamma \times M_{conv}$ for the maximum likelihood estimate (MLE) of the $\gamma \in \mathcal{R}^{1 \times N}$ coefficients, where we took the $\gamma$ estimates to be regional activity. We solved this MLE separately for each of the two conditions, providing distinct estimates of regional functional activation for 0-back and 2-back.

\subsubsection{Estimation of functional connectivity matrices}

\begin{figure}[H]
\begin{center}
\centerline{\includegraphics[]{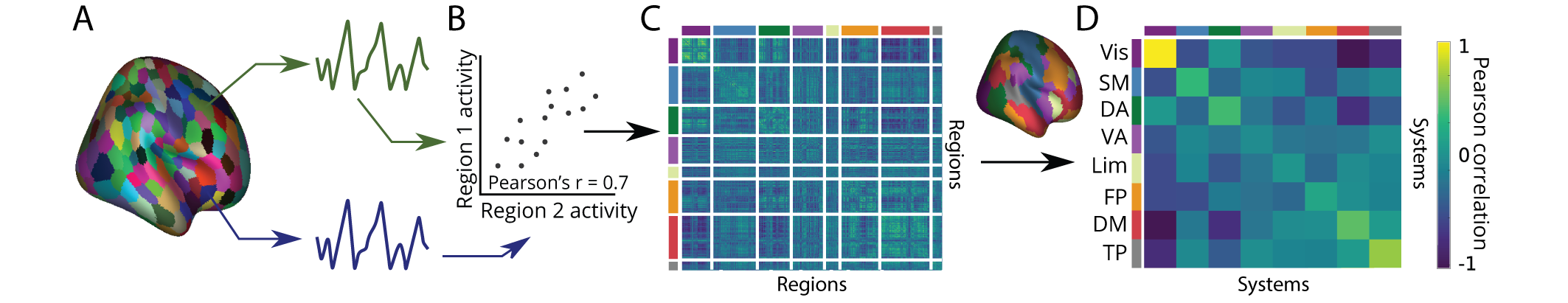}}
\caption{\textbf{Methodological schematic.} \emph{(A)} fMRI BOLD images from 644 subjects in the HCP S900 release were segmented into 400 regions to extract regional mean timeseries. \emph{(B)} We assessed the functional connectivity between each pair of regions by calculating the Pearson correlation coefficient between the time series of region $i$ and the time series of region $j$. \emph{(C)} We encoded all pairwise functional connectivity estimates in an adjacency matrix, which offers a formal representation of the network model under study. Each region was assigned to one of 8 intrinsic functional systems defined \emph{a priori}. \emph{(D)} From this assignment, we constructed a system-by-system functional connectivity matrix where each element indicates the average strength of all functional connections for region pairs in which one region of the pair is located in system $i$ and the other region of the pair is located in system $j$. Systems are color-coded and ordered from left to right (and from top to bottom) as follows: visual (Vis), somatomotor (SM), dorsal attention (DA), salience or ventral attention (VA), limbic (Lim), frontoparietal (FP), default mode (DM), and temporoparietal (TP).}\label{fig1}
\end{center}
\end{figure}

We used the preprocessed data to construct functional connectivity matrices reflecting functional interactions between regions and systems. Specifically, we extracted processed time series from each of the 400 regions in the Schaefer atlas (Fig. \ref{fig1} A). Next, we calculated the Pearson correlation coefficient between each pair of regional time series (Fig. \ref{fig1} B). We chose to use the Pearson correlation to represent functional connectivity due to its widespread use in the neuroimaging literature, as well as its ease of interpretability \cite{Zalesky2012a}, but we also demonstrate robustness of our results to other measures of functional connectivity in the Supplementary Materials. We collated all inter-regional estimates of functional connectivity into a single $400 \times 400$ connectivity matrix, $C_f$ (Fig. \ref{fig1} C), which we then treated as the formal encoding of a network model of brain function \cite{Bassett2018}. To be explicit, in this network model regions are represented by network nodes, and functional connections are represented by weighted edges, where the weight of the edge between node $i$ and node $j$ is given by the Pearson correlation coefficient between the time series of region $i$ and the time series of region $j$. Finally, we averaged the estimates of functional connectivity within systems, and between pairs of systems, to construct a system-by-system connectivity matrix (Fig. \ref{fig1} D).

\subsubsection{Identification of subnetworks using a weighted stochastic block model}

To partition the frontoparietal system into two functionally disjoint groups, we employed the weighted stochastic block model, which is a powerful community detection method. This method is complementary to the more widely used modularity maximization methods \cite{Newman2006}, but is noted to have increased flexibility and sensitivity to a more diverse set of network architectures \cite{Faskowitz2018,Betzel2018,Murphy2016,Betzel2018a,Betzel2018}. Briefly, the weighted stochastic block model is a generative model that places each of the $N$ nodes of network $C_f$ into one of \emph{K} communities. This placement is accomplished by finding a network partition $z \in \mathcal{Z}^{N\times 1}$ where $z_i \in \{1,2,\dots,K\}$ and $z_i$ denotes the membership of node \emph{i}. Assuming that network edge weights are normally distributed, following \cite{Aicher2015} and \cite{Betzel2018}, the generative model takes the following form:

\begin{equation}
P\left(C_f|z,\mu,\sigma^2 \right) = \prod_{i=1}^N \prod_{j=1}^N \exp\left(C_{f,ij}\cdot \frac{\mu_{z_iz_j}}{\sigma^2_{z_iz_j}}- \frac{C_{f,ij}^2 }{2\sigma^2_{z_iz_j}}-\frac{\mu^2_{z_iz_j}}{\sigma^2_{z_iz_j}}\right).
\end{equation}
\label{eq1}
Here we have introduced model parameters $\mu \in \mathcal{R}^{K\times K}$ and $\sigma^2 \in \mathcal{R}^{K\times K}$, where $\mu_{z_iz_j}$ and $\sigma^2_{z_iz_j}$ parameterize the weights of normally distributed connections between community $z_i$ and community $z_j$, and $C_{f,ij}$ denotes the $ij$th element of the network $C_f$. Furthermore, $P(C_f|z,\mu,\sigma^2)$ is the probability of generating the observed network $C_f$ given the parameters. This model is fit to $C_f$ in order to estimate the parameters $z$, $\mu$, and $\sigma^2$. We fit the model using MATLAB code provided in \cite{Aicher2015} and freely available at http://tuvalu.santafe.edu/\~\ aaronc/wsbm/. 

For each subject, we fit the weighted stochastic block model to the $m \times m$ subgraph of the adjacency matrix representing functional connections between the $m=61$ regions of the frontoparietal system. We selected $K=2$ \emph{a priori} due to prior evidence that the frontoparietal system can be separated into two distinct components \cite{Dixon2018}. The implementation generated a single maximum likelihood partition of regions into functional communities for each subject. Then, we pooled partitions across subjects and used a consensus similarity method \cite{Doron2012} to identify a single partition that is most similar to all others, where similarity is quantified by the $z$-score of the Rand coefficient \cite{Traud2011}. To assess the statistical significance of this partition, we performed a non-parametric permutation test. Specifically, for each subject, we calculated the log-likelihood of the weighted stochastic block model fitting the final consensus partition to their individual functional network. As a null, we calculated the log-likelihood of the weighted stochastic block model fitting a random permutation of the final consensus partition to their individual network. We assessed the difference between the true and null data using a multilevel model (see Section \ref{stats}).

\subsection{Analysis of diffusion tensor imaging data}

\subsubsection{Estimation of structural connectivity matrices}

After performing probabilistic tractography, we applied the same 400-region Schaefer atlas. Next, we calculated the proportion of streams seeded in a voxel in one region that reached another region. We chose to use the proportion of streamlines to represent structural connectivity due to the inhomogeneity of the region sizes. We collated all inter-regional estimates of structural connectivity into a single $400 \times 400$ connectivity matrix, $C_{s}$, which we then treated as the formal encoding of a network model of brain structure \cite{Bassett2018}. Similar to the model of brain function, in this structural network model, regions are represented by network nodes, and structural connections are represented by weighted edges, where the weight of the edge between node $i$ and node $j$ is given by the proportion of streams seeded at region $i$ that reach region $j$. Finally, we averaged the estimates of structural connectivity within systems, and between pairs of systems, to construct a system-by-system connectivity matrix, akin to the one that we constructed from the functional data.

\subsubsection{Calculation of boundary controllability}

We posited that structural connections between systems would play an important role in the functional coupling between the frontoparietal and default mode systems. Specifically, we hypothesized that the formal nature of that role was one of boundary controllability, more commonly studied in the field of control and dynamical systems theory \cite{Pasqualetti2014}. Boundary control is a quantifiable metric describing the notion that the topological location of a region within a \emph{structural} network partially governs that region's influence on the function of modules or communities in the network \cite{Betzel2016a,Gu2015a}. Intuitively, if region \emph{i} has strong structural connections to regions \emph{j} and \emph{l}, the activity of region \emph{i} influences the functional connection between regions \emph{j} and \emph{l}. Boundary control assesses the positioning of a region between two other regions, and can be calculated for region \emph{i} with respect to its control over regions \emph{j} and \emph{l}:
\begin{equation}
BC(i)=\left\{\begin{matrix}
1- \left( \frac{k_i(j)}{k_i}\right )^2- \left( \frac{k_i(l)}{k_i}\right )^2 & \textup{for}~k_i(j)+k_i(l)=k_i \\ 
\\
\left( \frac{k_i(j)}{k_i}\right )^2+ \left( \frac{k_i(l)}{k_i}\right )^2 & \textup{for}~k_i(j)+k_i(l)<k_i
\end{matrix}\right.
\label{eq2}
\end{equation}
Here, $k_i$, the degree of region \emph{i}, is the sum of all region \emph{i}'s structural connections. The variables $k_i(j)$ and $k_i(l)$ are the strength of region \emph{i}'s structural connections to regions \emph{j} and \emph{l}, respectively. A region with high boundary control is predicted to more effectively modulate the functional connection between region \emph{j} and region \emph{l}, although boundary control does not assess whether the modulation will increase or decrease the connection strength.

\section{Gene coexpression analysis}
\label{methods_coexpression}

To determine whether the two subnetworks that we identified in the frontoparietal system displayed distinct patterns of gene expression, we used gene expression data from six post-mortem brains available from the Allen Brain Institute. We focused our analyses on 16699 genes that had previously been identified as relevant for brain function \cite{Richiardi2015}. Data for these specific genes were available in 338 of the 400 brain regions. We assigned the anatomical location of each probe to one of 338 \emph{a priori} defined parcels. For each parcel and each gene, we calculated the mean expression of that gene across all probes, after subtracting the mean expression of each probe for that gene \cite{Krienen:2016e52}. Collectively, these calculations generated a data matrix of size 338 (parcels) by 16699 (mean expression across probes in that parcel for a given gene). 

Gene coexpression between parcel \emph{i} and parcel \emph{j} is measured by the Pearson correlation coefficient \emph{r} between gene expression values of parcel \emph{i} and gene expression values of parcel \emph{j}. To assess subnetwork specificity of gene coexpression, we first randomly sampled 500 genes and constructed a $338\times338$ gene coexpression matrix. Second, from this coexpression matrix, we calculated the mean gene coexpression both within and between subnetworks. Between subnetwork coexpression is the mean of the gene coexpression values for pairs of nodes for which one node in the pair is located in subnetwork (A) and the other node in the pair is located in subnetwork (B). Third, we calculated the ratio of within- to between-network gene coexpression, where a ratio $> 1$ indicates greater within-subnetwork coexpression than between-subnetwork coexpression. Finally, we compute the difference between this ratio and that expected in a nonparametric null model in which we randomly permuted subnetwork membership 1000 times, and recomputed coexpression ratios for each permutation. Next, we repeat the above process 10000 times, selecting a different random sample of 500 genes each time, generating a distribution of differences indicating whether the true ratio is larger than the mean of the null model ratios. To complete our statistical analysis, from this distribution we compute the probability of the difference being less than 0. We chose the approach of taking 10000 samples of 500 genes rather than taking a single sample of all the genes in order to demonstrate that the result is robust to the particular selection of genes, and to protect against our true ratio being influenced by outliers. For additional analyses controlling for the distance between regions, see the Supplemental Materials.

\subsection{Dynamical network model}
\label{s:neuralMass_methods}

As motivated more fully in the Results section, we propose that two subnetworks of the frontoparietal system work in functional opposition to either couple or decouple the frontoparietal system from the default mode system. Specifically, we suggest that the functional connection between the frontoparietal and default mode systems is governed by the relative activation amplitudes of the two frontoparietal subnetworks. To further probe these relationships and their mechanistic underpinnings, we built a coarse-grained network in which each unit represented a particular brain system, and we simulated system dynamics with a canonical coupled oscillator model. More specifically, network activity was modeled by the normal form of a supercritical Hopf bifurcation (also referred to as the Stuart-Landau model), which describes the transition between a state of low activity and a state of oscillatory dynamics \cite{hoppensteadt2012weakly,kuznetsov2013elements}. We chose this model because (i) it permits the independent manipulation of oscillator amplitudes (Fig. S1), allowing us to further investigate the empirically observed relationships between activity and connectivity, and (ii) it is often used to model large-scale brain activity \cite{Freyer2011,Freyer2012,Moon2015,Deco2017,Senden2017}. 
Following \cite{Deco2017,Senden2017}, the local dynamics of the $j^{th}$ unit are given by the following equation:
\begin{equation}
	\frac{du_{j}}{dt} = u_{j}[a_{j} + i\omega_{j} - |u_{j}|^{2}] + \xi\eta_{j}(t),	
\label{eq:hopf_complex_local}
\end{equation}
\noindent where $u_{j} = \rho_{j}e^{i\theta_{j}} = x_{j} + iy_{j}$, $\eta_{j}$ is drawn from a normal distribution to add Gaussian noise to the system, and $\xi$ scales the noise. In Eq.\eqref{eq:hopf_complex_local}, the term $a_{j}$ is commonly called the \emph{bifurcation parameter}. When $a_{j} < 0$, the system goes to a low-activity fixed point and when $a_{j} > 0$, the system obeys a stable limit-cycle solution with angular frequency $\omega_{j}$ and signal amplitude governed by $a_{j}$.

Following \cite{Deco2017,Senden2017}, we model a network of interacting components by separating the system into its real and imaginary parts, and we link different components via the following set of coupled differential equations:

\begin{equation}
\frac{dx_j}{dt}=\left[a_j - x_j^2 - y_j^2\right]x_j - \omega_jy_j+G\sum_{i=1}^{n_o} D_{ij} \left(x_i-x_j \right) + \xi \eta_j(t),
\label{eq4}
\end{equation}

and 

\begin{equation}
\frac{dy_j}{dt}=\left[a_j - x_j^2 - y_j^2\right]y_j + \omega_jx_j+G\sum_{i=1}^{n_o} D_{ij} \left(y_i-y_j \right) + \xi \eta_j(t).
\label{eq5}
\end{equation}

\noindent Here, \emph{G} is the coupling strength and $n_o$ is the total number of oscillators, which in our case is four (default mode network, dorsal attention network, frontoparietal subnetwork (A), and frontoparietal subnetwork (B)). As suggested by \cite{Senden2017} we set $\xi = 0.02$, and we took $x_{j}$ as the oscillatory signal of interest. To estimate the frequency parameters, we empirically calculated the peak frequency of each of the 4 systems during the resting state, fit a normal distribution to the peak frequency values, and drew from the normal distribution. Similarly, to establish $D_{ij}$, the coupling of the network nodes, we calculated the mean structural connectivity estimated from diffusion tractography between system $i$ and system $j$ across all subjects. We integrated the equations using a time step of 0.01 seconds for 6 minutes, which was the approximate length of the task scans. 

To estimate reasonable values for the coupling and bifurcation parameters, we conducted a parameter sweep (Fig. S2), and computed both the root mean square of the time series, and the synchrony among all oscillators using the Kuramoto order parameter, defined at time point \emph{t} as:
\begin{equation}
R(t) = \frac{1}{n_o}\left|\sum_{j=1}^{n_o} \textup{e}^{i \phi_j(t)} \right|,
\label{eq5}
\end{equation}
where $\phi_j(t)$ is the instantaneous phase of oscillator \emph{j} at time \emph{t}. To get a summary statistic for the entire time series, we took the mean of \emph{R} across time (Fig. S2 E). The instantaneous phase was computed by taking the Hilbert transform of the unfiltered time-series.

For our baseline working point, we selected $a = -0.075$ and $G = 0.1$ for all units, where the Kuramoto order parameter has an intermediate value, signifying a realistic dynamical regime between a state of no synchrony and a state of complete synchrony among all oscillators. Furthermore, at this working point the root mean square of the time series is higher than the noise level (Fig. S2 F). For each point in our grid defined by $a$ and $G$, we averaged results from 25 initializations of the system. Importantly, the bifurcation parameter of a node in this four oscillator system is linearly related to the root mean square of the time series values of that node (Fig. S1 B).

\subsection{Statistical analysis}
\label{stats}
At several points, we calculate the statistical difference between outcome variables of the two sub-networks. To that end, we initially take a parametric approach, and then we confirm all of our findings using a non-parametric permutation-based approach. In all visualizations of statistical relationships, subject effects have been regressed out from the dependent variable.

\subsubsection{Subnetwork connection differences}
In testing our hypotheses, we often asked questions of the following form: Does the strength of the connection between subnetwork (A) and the default mode system differ from the strength of the connection between subnetwork (B) and the default mode system? For questions of this form, we used a multilevel model where each outcome variable (e.g., a measurement of connection strength) has attributes encoding subnetwork membership, task run, and subject identity. The multilevel model framework \cite{tom1999multilevel} accounts for the nested nature of the data (multiple scans nested within subject). We specified the model as:
\begin{equation}
\mathrm{OutcomeVariable}_{it} = \mathcal{B}_{0i} + \mathcal{B}_{1i}\mathrm{SubNetwork}_{it}+e_{it},
\end{equation}
where $\mathrm{OutcomeVariable}_{it}$ is the outcome variable (i.e. connection strength) for person $i$ on run $t$;  $\mathcal{B}_{0i}$ indicates the level of the outcome in subnetwork (A); $\mathcal{B}_{1i}$ indicates differences in the level of outcome associated with subnetwork (B) versus subnetwork (A); and $e_{it}$ are residuals. 

Person-specific intercepts (from Level 1) were specified (at Level 2) as:
\begin{equation}
\mathcal{B}_{0i} = \gamma_{00} + u_{0i},
\end{equation}
and
\begin{equation}
\mathcal{B}_{1i} = \gamma_{10},
\end{equation}
where $\gamma$ denotes a sample-level parameter and $u$ denotes residual between-person differences that may be correlated, but are uncorrelated with $e_{it}$. The multilevel model was fit with \emph{lme} in R using maximum likelihood estimation. In the case of many outliers, we treat our data with robust models, rather than standard linear models. Robust models down-weight points of data, where the most outlying points are down-weighted most severely. Specifically, we implement robust multilevel models using robustlmm in R \cite{koller2016robustlmm}. We note in the text whenever a robust multilevel model is used.

\subsubsection{Repeated mesures correlations}
Unless otherwise noted, we use a repeated measures correlation when examining the association between two continuous variables \cite{bakdash2017repeated}. The repeated measures correlation accounts for non-independence among observations (due to multiple runs per subject) by using a form of analysis of covariance (ANCOVA) to adjust for between-person variance. The model is specified as:
\begin{equation}
\mathrm{Measure1}_{it} = \overline{\mathrm{Measure2}}_{i} + \mathrm{Subject}_i + c(\mathrm{Measure2}_i) + e_{it},
\end{equation}
where $\mathrm{Measure1}_{it}$ is the value of variable one for subject $i$ during measurement occasion $t$, $\overline{\mathrm{Measure2}}_{i}$ is the mean value of the second variable in the $i$-th participant, $\mathrm{Subject}_i$ is a unique identifier for each participant, and $c(\mathrm{Measure2}_i)$ is the covariate for the $i$-th participant and is equal to $\mathcal{B}(\mathrm{Measure2}_{it} - \overline{\mathrm{Measure2}}_i)$, where $\mathcal{B}$ is the slope coefficient of the covariate. Like a Pearson correlation coefficient ($r$), the repeated measures correlation ($rrm$) is bounded by -1 to 1, and represents the strength of the linear association between two variables. The repeated measures correlation was estimated using the rmcorr package in R \cite{bakdash2017repeated}.

\subsubsection{Nonparametric permutation-based approach}
In addition to the multilevel linear model, we employ a complementary permutation-based approach. We begin with vectors $Y_1\in \mathcal{R}^{1\times4n}$, $Y_2\in \mathcal{R}^{1\times4n}$, and $S\in \mathcal{R}^{1\times4n}$, and we wish to test whether there is a difference in the means of $Y_1$ and $Y_2$ against a null model. To construct the null model, for subject $i$ we find the two entries $J=(j_1, j_2)$ for which $S = i$. We then randomly reassign elements $J$ between $Y_1$ and $Y_2$, and repeat this procedure for all subjects to construct a null $Y^{null}_1$ and null $Y^{null}_2$ where we would expect the means to be equal. We then calculate the mean difference $d_{null} = \textup{mean}(Y^{null}_1-Y^{null}_2)$. We re-permute and recalculate the mean 10000 times to establish a null distribution of the difference, and we determine a $p$-value for the true effect by calculating the proportion of null differences that are greater than the observed difference.

\section{Results}

\subsection{Connectivity between default mode and frontoparietal systems tracks working memory performance.}

\begin{figure}[H]
\centerline{\includegraphics[]{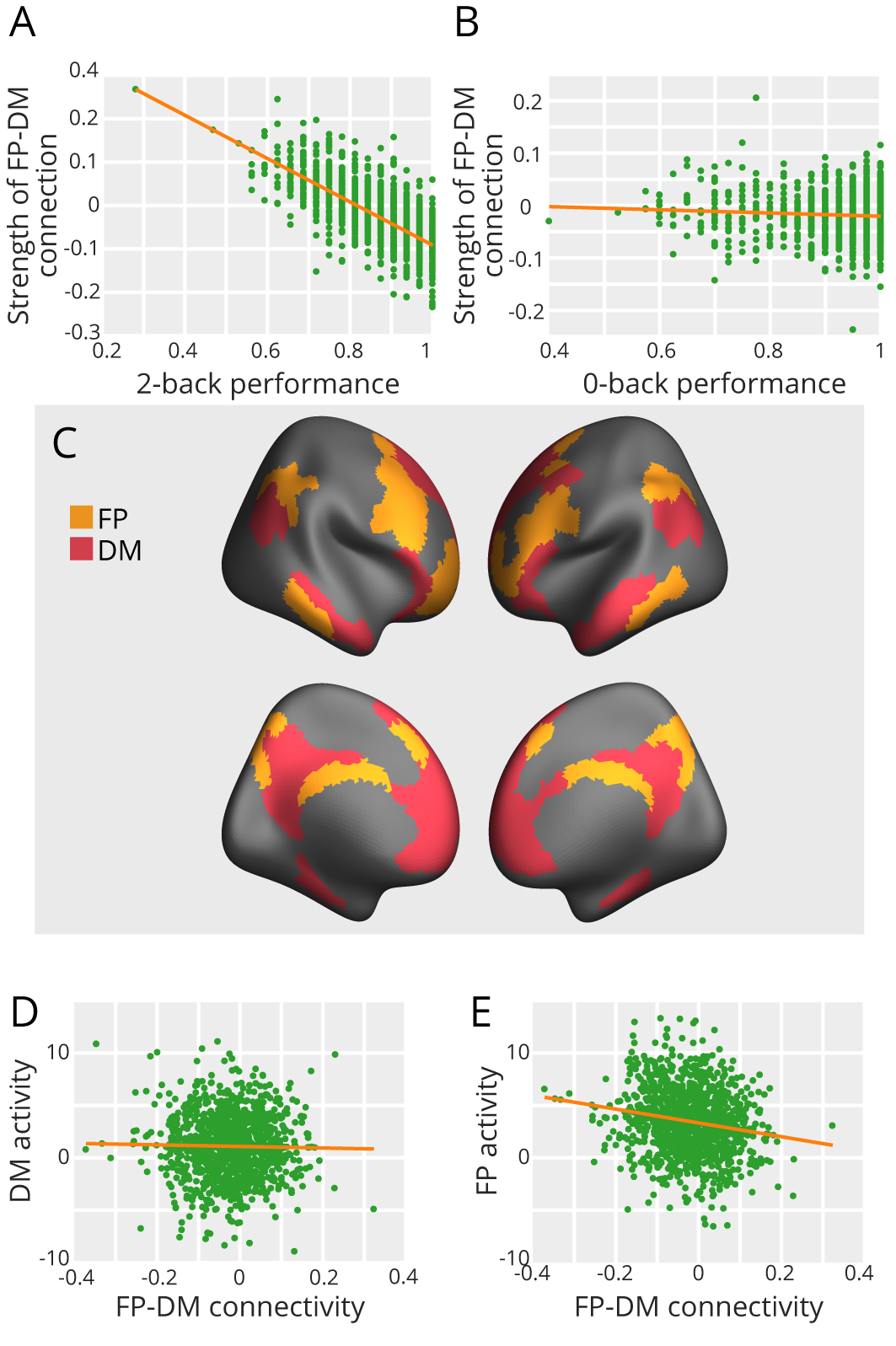}}
\caption{\textbf{Functional connectivity and activity in the frontoparietal system relate to working memory performance.} \emph{(A)} We found that the strength of the connection between the frontoparietal and default mode systems is negatively correlated with performance on the 2-back working memory task (repeated measures correlation, $r = -0.4731$, $p < 0.001$, $DF = 558$. \emph{(B)} During the 0-back task, we found no significant relationship between performance and the strength of the connection between the frontoparietal and default mode systems (repeated measures correlation; $r = -0.0293$, $p = 0.487$, $DF = 562$). \emph{(C)} Anatomical location of regions within the frontoparietal and default mode systems, displayed on the cortical surface. \emph{(D)} Activity of the default mode system was not significantly correlated with the strength of the functional connection between the frontoparietal and default mode systems (repeated measures correlation; $r = -0.0441$, $p = 0.296$, $DF = 562$). \emph{(E)} Activity of the frontoparietal system was negatively correlated with the strength of the functional connection between the frontoparietal and default mode systems (repeated measures correlation; $r = -0.0859$, $p = 0.042$, $DF = 562$).}\label{fig2}
\end{figure}

We first sought to determine whether the strength of the connection between the default mode and frontoparietal systems tracks performance on a 2-back working memory task. For each participant, we calculated the average functional connectivity between all relevant pairs of brain regions, where one region of the pair was located in the default mode system and the other region of the pair was located in the frontoparietal system. We then estimated the relationship between behavioral accuracy and between-system strength. We found that the strength of the between-system connection was negatively correlated with performance on the task across subjects (repeated measures correlation, $r = -0.4731$, $p < 0.001$, $DF = 558$; Fig.~\ref{fig2}A). 

To assess task-specificity, we repeated the analysis using the imaging and performance data from the 0-back condition. During this condition, minimal working memory load exists. We found no significant relationship between behavioral performance and the strength of the connection between the default mode and frontoparietal systems (repeated measures correlation; $r = -0.0293$, $p = 0.487$, $DF = 562$; Fig.~\ref{fig2}B-C). To assess the robustness of the connectivity-behavior relationship, we used an alternate behavioral measure (see Supplementary note 1 and Fig. S4 ), an alternate network parcellation (see Supplementary note 2 and Figs. S5, S6), and an alternate measure of functional connectivity (see Supplementary note 3 and Fig. S7). The results of these additional analyses serve to confirm our main findings.

\subsection{The activity of the frontoparietal system is correlated with the strength of the functional connection between the frontoparietal and default mode systems.}

After observing a statistical relation between working memory performance and the functional connectivity between the default mode and frontoparietal systems, we sought to better understand its potential drivers. We began by testing the hypothesis that system activity drives inter-system connectivity. We defined regional activity as the $\beta$-weight from a GLM fit to the regional BOLD magnitude during the 2-back task. To assess the activity of a system, we averaged those GLM $\beta$-weights across all regions in the frontoparietal system or across all regions in the default mode system. We observed that the frontoparietal system was more active than the default mode system (frontoparietal: mean $= 2.74$, 95\% CI [2.51, 2.98]; default mode: mean $= -3.09$, 95\% CI [-3.33, -2.85]). We confirmed this difference in activity between the two systems using a multilevel model ($\beta = -7.4501$, $p < 0.0001$, $t(1769) = -35.7$, $SE =  0.20847$, $n = 2414$), and we found similar results using a non-parametric permutation test ($p<0.0001$); see Section \ref{stats}.

In assessing the relationship between system activity and inter-system connectivity, we found that the activity of the default mode system is unrelated to the strength of the connection between the default mode and frontoparietal systems (repeated measures correlation; $r = -0.0441$, $p = 0.296$, $DF = 562$; Fig.~\ref{fig2}D). In contrast, we found that the activity of the frontoparietal system was negatively correlated with the strength of this connection (repeated measures correlation; $r = -0.0859$, $p = 0.042$, $DF = 562$; Fig.~\ref{fig2}E). Given that frontoparietal activity is correlated with the strength of the inter-system connectivity, and that the inter-system connectivity is correlated with behavioral performance, it stands to reason that the frontoparietal activity should also be correlated with behavioral performance. Indeed we found that the activity of the frontoparietal system is positively related to behavioral performance (repeated measures correlation; $r =  0.1019$, $p = 0.016$, $DF = 558$; Fig. S3 A), and this relation was not observed when considering activity of the default mode system (repeated measures correlation; $r =  0.0384$, $p = 0.363$, $DF = 558$; Fig. S3 B).

\subsection{Subnetworks of the frontoparietal system differentially modulate the strength of the functional connection between the frontoparietal and default mode systems.}

\begin{figure}
\centerline{\includegraphics[]{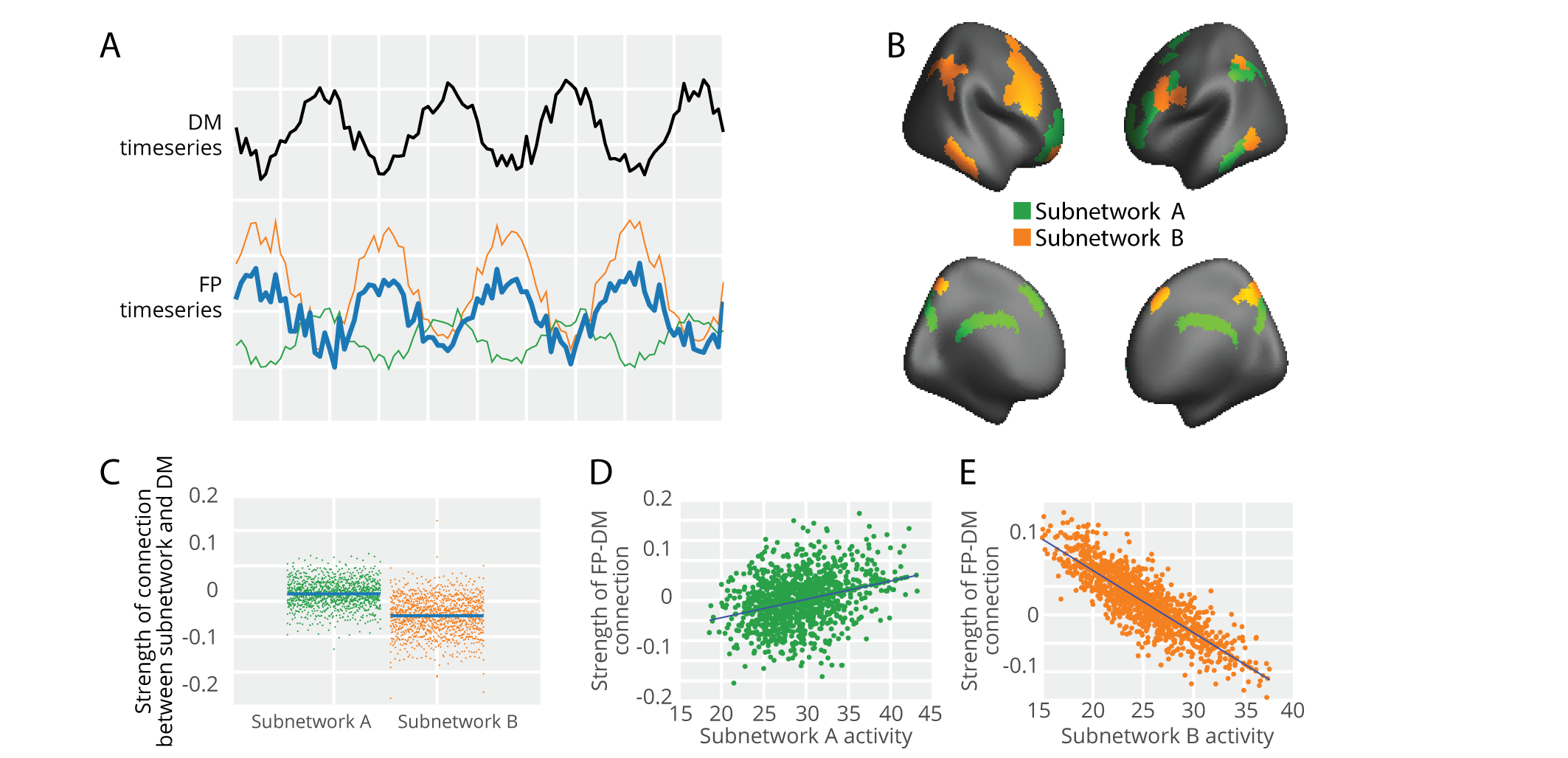}}
\caption{\textbf{Subnetworks of the frontoparietal system.} \emph{(A)} We hypothesized that the strength of the connection between the frontoparietal and default mode systems can be tuned by altering the relative amplitudes of subnetworks within the frontoparietal system. \emph{(B)} Community detection reveals two distinct frontoparietal subnetworks, which we show here projected onto the cortical surface. \emph{(C)} We found that the activity of subnetwork (A) was positively correlated with the activity of the default mode system (mean $r= 0.042$, 95\%CI: [0.038, 0.046]), while the activity of subnetwork (B) was negatively correlated with the activity of the default mode system (mean $r= -0.082$, 95\%CI: [-0.088, -0.076]). \emph{(D)} Using a simple regression model, we tested whether the strength of the functional connection between the default mode and frontoparietal systems could be predicted by a linear combination of the activity of subnetwork (A) and the activity of subnetwork (B). Within this model, we found that an increase in subnetwork (A) activity corresponds to an increase in the strength of the functional connection between the frontoparietal and default mode systems (estimate of regression coefficient $\beta = 0.006535$, 95\% CI: (0.00519, 0.00788)). \emph{(E)} Conversely, within the same model we found that an increase in subnetwork (B) activity corresponds to a decrease in the strength of the functional connection between the two systems (estimate of regression coefficient $\beta = -0.0112$, 95\% CI: (-0.01273, -0.0097)). }\label{fig3}
\end{figure}

The results reported in the previous section are correlative, and to press further we must posit a testable causal model. We propose that the dynamics of the frontoparietal system directly modulate the strength of the functional connection between the frontoparietal and default mode systems. Our candidate mechanism assumes that the frontoparietal system is composed of two distinct, non-overlapping sub-systems. Subnetwork (A) is posited to display dynamics that are correlated with the dynamics of the default mode system, while subnetwork (B) is posited to display dynamics that are anticorrelated with the dynamics of the default mode system (Fig.~\ref{fig3}A). Further, we propose that the relative activity magnitudes of these two systems interact to tune the strength of the inter-system (frontoparietal - default mode) connection. When subnetwork (A) is highly active relative to subnetwork (B), the strength of the inter-system connection will be positive; conversely, when subnetwork (B) is highly active relative to subnetwork (A), the strength of the inter-system connection will be negative. 

To assess the validity of this conceptual model, we first test the assumption that the frontoparietal system is composed of two distinct, spatially non-overlapping sub-systems. For each subject, we applied a weighted stochastic block model with $K=2$ to the subgraph of the adjacency matrix representing functional connections between frontoparietal regions, and we then extracted a group-representative partition using a consensus similarity method \cite{Doron2012} (see Methods). The two subnetworks are shown on the cortex in Fig.~\ref{fig3}B. To assess the statistical significance of this partition, we performed a non-parametric test, permuting the association of regions to subnetworks (see Methods). Against the null model, we found that this final consensus partition had a significantly higher log-likelihood using a multilevel model ($\beta = -492.57$, $p < 0.0001$, $t(1769) = -71.0$, $SE = 6.9371$, $n = 2414$), and we found similar results using a non-parametric permutation test ($p<0.0001$).

To further validate their biological distinctness, we sought to determine whether the two subnetworks showed distinguishable patterns of gene expression. To this end, we quantified the average magnitude of gene coexpression for pairs of regions for which both regions were located \emph{within} a single subnetwork. We also quantified the average magnitude of gene coexpression for pairs of regions for which one region of the pair was located in one subnetwork and the other region of the pair was located in the other subnetwork (see Methods). When tested against a non-parametric null model, we found that gene coexpression \emph{within} subnetwork (A) was significantly higher than gene coexpression \emph{between} subnetwork (A) and subnetwork (B) ($p = 0.0057$). Similarly we found that gene coexpression \emph{within} subnetwork (B) was higher than gene coexpression \emph{between} subnetwork (B) and subnetwork (A) ($p=0.0111$). For complementary analyses demonstrating that this result cannot be explained by distances between regions, see Supplementary note 4. These findings support the notion that subnetwork (A) and subnetwork (B) are biologically distinct sectors of the frontoparietal system.

Next, we sought to test our proposition that -- of the two frontoparietal subnetworks -- one displays dynamics that are correlated with the dynamics of the default mode system, while the other displays dynamics that are anticorrelated with those of the default mode system. For each subject and each subnetwork, we calculated the strength of the functional connection between all pairs of regions for which one region of the pair is located in the subnetwork and the other region of the pair is located in the default mode system. Consistent with our hypothesis, we found that the activity of subnetwork (A) is positively correlated with the activity of the default mode system (mean $r= 0.042$, 95\%~CI: [0.038, 0.046]; Fig.~\ref{fig3}B left), while the activity of subnetwork (B) is negatively correlated with the activity of the default mode system (mean $r= -0.082$, 95\%~CI: [-0.088, -0.076]; Fig.~\ref{fig3}B right). Using a multilevel model, we found that these correlations are significantly different ($\beta = -0.12469$, $p<0.0001$, $t(1769) = -37.4$, $SE = 0.0033272$, $n = 2414$), and we found similar results using a non-parametric permutation test ($p<0.0001$). To further unpack these findings, we also considered the relation between these two subnetworks and the dorsal attention system, which has been described as antagonistic to the default mode system in working memory tasks \cite{Anticevic2012,Spreng2010,Dixon2017}. Consistent with this account, we found that the activity of subnetwork (A) was negatively correlated with the activity of the dorsal attention system (mean $r= -0.14$, 95\%~CI: [-0.18, -0.11]; Fig. S8 A right), while the activity of subnetwork (B) was positively correlated with the activity of the dorsal attention system (mean $r= 0.12$, 95\%~CI: [0.11, 0.12]; Fig. S8 A left). Using a multilevel model, we found that these correlations are significantly different ($\beta = 0.13057$, $p<0.0001$, $t(1769) = 46.8$, $SE =  0.0027885$, $n = 2414$), and we found similar results using a non-parametric permutation test ($p<0.0001$).

To address the final proposition in our model, we sought to determine whether increased subnetwork (A) activity would lead to a stronger positive functional connection between the frontoparietal and default mode systems, while increased subnetwork (B) activity would lead to a stronger negative functional connection between the two systems. Because the summation of the subnetwork timeseries should reflect the complete frontoparietal timeseries, we also reasoned that when subnetwork (A) is more active (higher amplitude) relative to subnetwork (B), the frontoparietal signal would be more similar to the dynamics of subnetwork (A) than to the dynamics of subnetwork (B). To test these expectations, we began by quantifying subnetwork activity using the root mean square (RMS) of the subnetwork timeseries. We then used a single robust linear model to explain the strength of the functional connection between the default mode and frontoparietal systems by a linear combination of the activity of subnetwork (A) and the activity of subnetwork (B). Consistent with our hypothesis, we found that an increase in subnetwork (A) activity corresponded to a stronger positive functional connection between the two systems (estimate of regression coefficient $\beta = 0.006535$, 95\% CI: (0.00519, 0.00788)); Fig. \ref{fig3}D), while an increase in subnetwork (B) activity corresponded to a stronger negative functional connection between the two systems (estimate of regression coefficient $\beta = -0.0112$, 95\% CI: (-0.01273, -0.0097)); Fig.~\ref{fig3}E). In a complementary analysis, we used a single robust linear model to explain the strength of the functional connection between the \emph{dorsal attention system} and the frontoparietal system by a linear combination of the activity of subnetwork (A) and the activity of subnetwork (B). We found that an increase in subnetwork (A) activity corresponded to a stronger \emph{negative} functional connection between the two systems ($\beta = -0.00553$, 95\% CI: (-0.00677, -0.00486)), while an increase in subnetwork (B) activity corresponded to a stronger \emph{positive} functional connection between the two systems ($\beta =  0.00989$, 95\% CI: (0.00848, 0.011316)).

The results of the three tests described above serve to validate the formal structure of our model. Next we turned to an assessment of the relevance of this complex dynamical system for behavior. Specifically, we had observed previously that behavioral performance decreases as the correlation between the frontoparietal and default mode systems increases. Here we seek to explain that observation using the activity of the two subnetworks. Because subnetwork (A) is positively related to frontoparietal-default mode connectivity, which is itself negatively related to behavioral performance, we would expect subnetwork (A) activity to be negatively related to behavioral performance. Conversely, because subnetwork (B) is related negatively to frontoparietal-default mode connectivity, which is itself negatively related to behavioral performance, we would expect subnetwork (B) activity to be positively related to behavioral performance. To probe these relationships, we fit a single robust multilevel model with behavioral performance as the dependent variable, and subnetwork (A) activity and subnetwork (B) activity as independent variables. As expected, we found that subnetwork (A) activity is negatively related to behavioral performance ($\beta =  -0.00271$, 95\% CI: (-0.0041, -0.00132)), and subnetwork (B) activity is positively related to behavioral performance ($\beta = 0.00273$, 95\% CI: (0.00115, 0.00430)).

\subsection{The structural role of the frontoparietal subnetworks}

We next turn to an examination of what, if any, neuroanatomical support exists for the subnetwork-driven dynamics espoused in the previous section. Recent advances in network control theory have posited that changes in the activation of single brain regions can induce a propagation of activity along white matter tracts to affect distributed circuit behavior in a predictable fashion \cite{Gu2015a,Tang2018}. Here we test this notion within the specific confines of our experiment, asking: Does the structural connectivity of frontoparietal subnetworks constrain how activity propagates to neighboring areas, thereby modulating the coupling between the frontoparietal and default mode systems? We hypothesize that subnetwork (A) is more structurally connected to the default mode system, while subnetwork (B) is more structurally connected to the dorsal attention system. This hypothesis is based on evidence that a higher number of white matter tracts between two regions can support stronger functional connectivity between them \cite{Skudlarski2008}, allowing subnetwork (A) to strongly couple to the default mode system and allowing subnetwork (B) to strongly couple to the dorsal attention system. 
 
\begin{figure}[H]
\centerline{\includegraphics[]{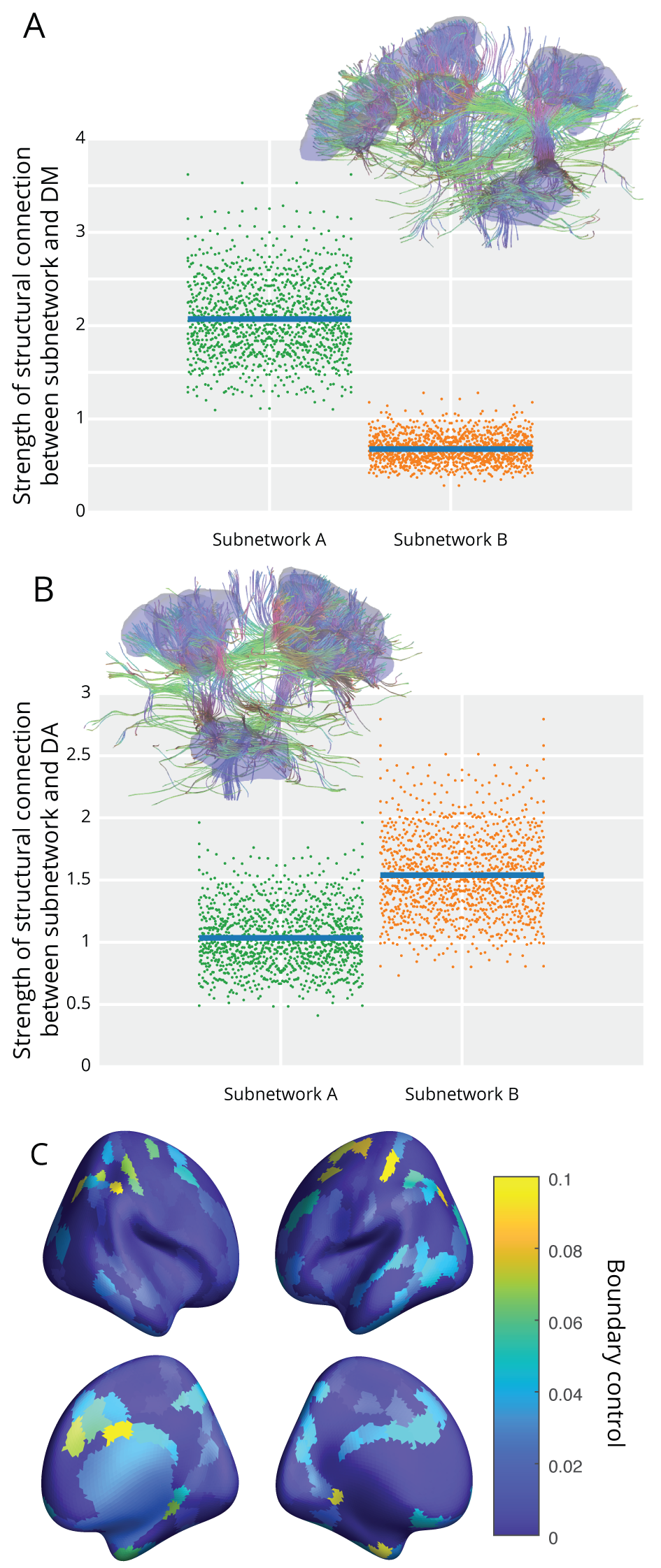}}
\caption{\textbf{White matter connectivity of the frontoparietal subnetworks.} \emph{(A)} Subnetwork (A) is more strongly structurally connected to the default mode system (mean = 2.07, 95\%~CI: (2.05, 2.09)) than is subnetwork (B) (mean = 0.676, 95\%~CI: (0.668, 0.685)). \emph{(B)} Subnetwork (A) is less strongly structurally connected to the dorsal attention system (mean = 1.03, 95\%CI: (1.02, 1.04)) than is subnetwork (B) (mean = 1.54, 95\%~CI: (1.52, 1.56)). Note that for visualization purposes, the data was not visually adjusted to account for two points coming from each subject, whereas the reported statistics do take this into account. The insets of panels \emph{(A)} and \emph{(B)} display the white matter fibers emanating from subnetworks (A) and (B), respectively. \emph{(C)} The anatomical distribution of boundary control calculated with respect to the default mode and dorsal attention systems is over represented in the frontoparietal system in comparison to a non-parametric permutation-based null model.}\label{fig4}
\end{figure}

We tested this hypothesis by calculating the strength of the structural connectivity between subnetworks and systems using diffusion imaging tractography (see Methods). Consistent with our hypothesis, we found that subnetwork (A) is more strongly connected to the default mode system (mean = 2.07, 95 \%CI: (2.05, 2.08); Fig. \ref{fig4}A left) than is subnetwork (B) (mean = 0.676, 95\%CI: (0.668, 0.685); Fig. \ref{fig4}A right). Using a multilevel model, we found that these correlations were significantly different ($\beta = -0.0013939$, $p<0.0001$, $t(1769) = -174.8$, $SE = 7.9759\times 10^{-6}$, $n = 2414$), and we found similar results using a non-parametric permutation test ($p<0.0001$). Similarly, we found that subnetwork (B) is more strongly connected to the dorsal attention system (mean = 1.54, 95\% CI: (1.52, 1.56); Fig. \ref{fig4}B left) than is subnetwork (A) (mean = 1.03, 95\% CI: (1.02, 1.04); Fig. \ref{fig4}B right). Using a multilevel model, we found that these correlations were significantly different ($\beta = 0.00050413$, $p<0.0001$, $t(1769) = 99.3$, $SE = 5.0721\times 10^{-6}$, $n = 2414$), and we found similar results using a non-parametric permutation test ($p<0.0001$). Further, across subjects we found that the less structurally connected the two subnetworks were, the better individuals tended to perform on the 2-back working memory task ($r = - 0.0766$, $p = 0.03$). 

Collectively, these data suggest that the frontoparietal subnetworks might be optimally positioned in the structural connectome to mediate coupling between the default mode and dorsal attention areas, a coupling that is negatively correlated with performance (Fig. S12). To more directly test this notion, we calculated the regional boundary control (Eq. \ref{eq2}) with respect to the default mode and dorsal attention systems (Fig.~\ref{fig4}C). Of the 20 regions for which boundary control exceeded the 95-th percentile, 9 were located in the frontoparietal system. To evaluate statistical significance, we compared these results to those of a non-parametric null model, in which we randomly permute the association between boundary control values and brain regions. We found that the probability that 9 or more of the top 20 regions fell within the frontoparietal system was significant with respect to the null model ($p = 0.0019$). In summary, the data suggest that the frontoparietal system is structurally positioned to effectively control the coupling between the default mode and dorsal attention systems. 

\subsection{A reduced dynamical model for probing the relation between system activity and connectivity.}

\begin{figure}
\centerline{\includegraphics[]{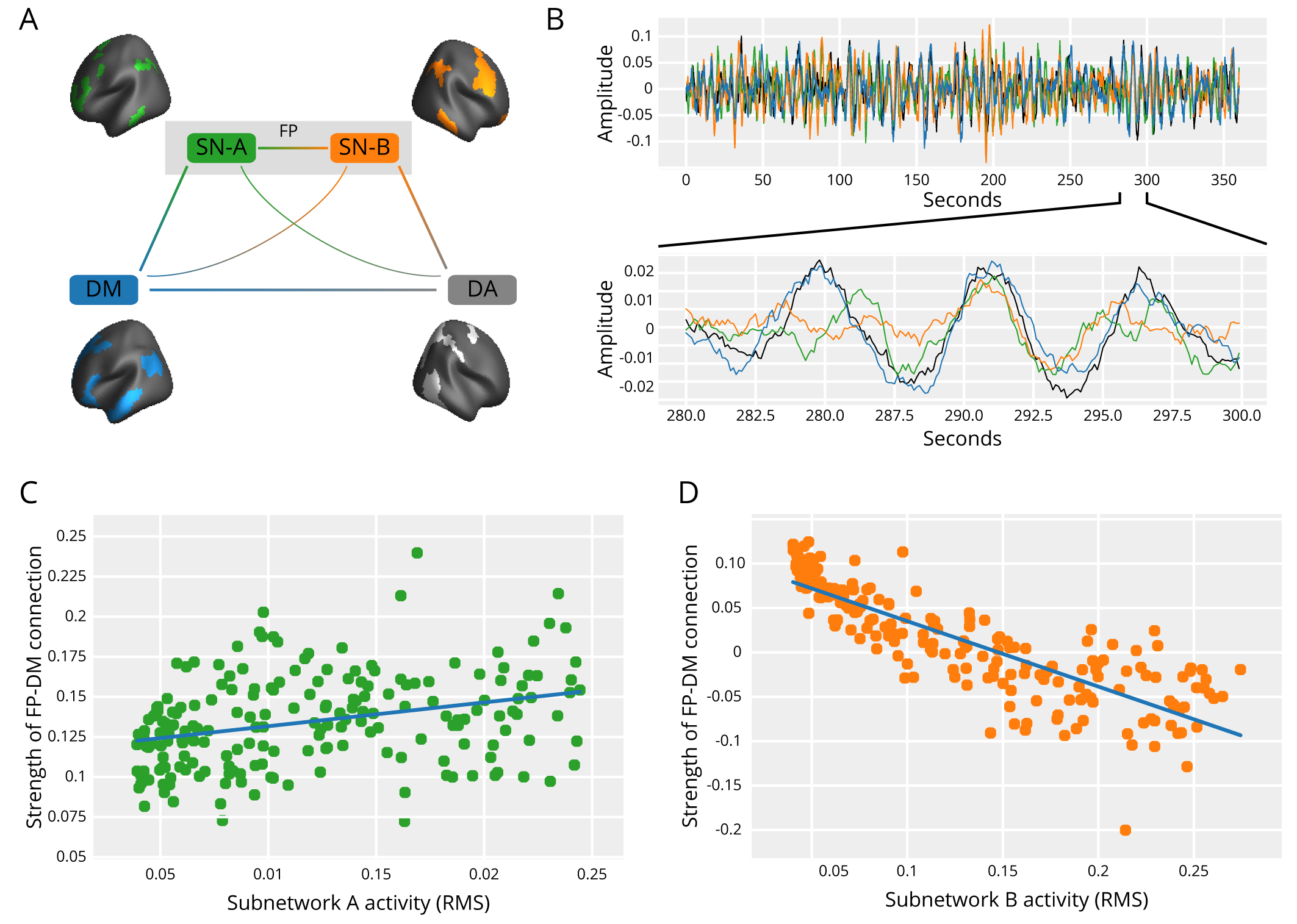}}
\caption{\textbf{A simplified model for studying the relationships between brain system activity and connectivity.} \emph{(A)} We constructed a dynamical model of a 4-component network consisting of the frontoparietal subnetwork (A), the frontoparietal subnetwork (B), the default mode system, and the dorsal attention system. Oscillatory activity of each unit was modeled using the normal form of a Hopf bifurcation and the relative coupling strength between systems was determined empirically from diffusion tensor imaging data. \emph{(B)} The network dynamics were integrated out to a total time of 6 minutes, consistent with the length of the empirical n-back scan. \emph{(C)} Increasing the amplitude of subnetwork (A) activity, while keeping all others equal, caused an increase in the functional connectivity between the frontoparietal and default mode units (Pearson correlation coefficient between unit time series: $r^2 = 0.157$, $p=0.001$). All calculations were performed after subtracting the mean network connectivity. \emph{(D)} Increasing the amplitude of subnetwork (B) activity caused a decrease in the functional connectivity between the frontoparietal and default mode units (Pearson correlation coefficient between unit time series: $r^2 = 0.75$, $p<0.0001$).}
\label{fig6} 
\end{figure}

In the previous sections, we found evidence consistent with (but not proving) the causal notion that increased activity of subnetwork (A) drives stronger coupling between the frontoparietal and default mode systems, while increased activity of subnetwork (B) drives anticorrelation between the two systems. While it is difficult to prove the validity of such a causal model in healthy human participants, we can gather additional supportive evidence from \emph{in silico} experiments exercising a formal computational model of the dynamical system. Specifically, we modeled the dynamics of a reduced four node network, with nodes representing the frontoparietal subnetwork (A), the frontoparietal subnetwork (B), the default mode system, and the dorsal attention system (Fig.~\ref{fig6}A). We consider each unit in the network to be an oscillator, with dynamics described by the normal form of a Hopf bifurcation and with frequencies randomly sampled from an empirically measured distribution (see Sec.~\ref{s:neuralMass_methods} for details). We coupled the four systems according to the mean weight of the structural connections between them, as estimated from diffusion tensor imaging tractography, averaged across subjects (see Methods). The model has two free parameters: (i) the global coupling parameter, which tunes the general capacity for synchronization, and (ii) the bifurcation parameter of each oscillator, which tunes the amplitude of the oscillator time series (Fig. S1). Following a broad parameter sweep, we selected parameter values to ensure a realistic dynamical regime between a state of no synchrony and a state of complete synchrony among all oscillators (see Methods and Fig.~\ref{fig6}B).

Next, we implemented the model to probe the relation between subnetwork activity and connectivity, focusing initially on the connectivity between the frontoparietal and default mode systems. In agreement with our empirical results, we found that increasing the amplitude of subnetwork (A) activity increased the correlation between the frontoparietal and default mode unit time series (Pearson's correlation coefficient $r^2 = 0.157$, $p=0.001$; Fig.~\ref{fig6}C). Similarly, and again in agreement with our empirical results, we found that increasing the amplitude of subnetwork (B) activity decreased the correlation between the frontoparietal and default mode unit timeseries (Pearson's correlation coefficient $r^2 = 0.75$, $p<0.0001$; Fig.~\ref{fig6}D). To assess the reliability of these results, we performed the same numerical experiments for a range of coupling and bifurcation parameter values, across which the effects remained robust (Fig. S10). Next, we examined the complementary relation between subnetwork activity and the connectivity between the frontoparietal and dorsal attention systems. Again, consistent with our empirical results, we found that increasing subnetwork (A) activity decreased the correlation between the frontoparietal and dorsal attention unit timeseries (Pearson's correlation coefficient $r^2 = 0.65$, $p<0.0001$; Fig. S11 A), and increasing subnetwork (B) activity increased the correlation between the frontoparietal and dorsal attention unit time series (Pearson's correlation coefficient $r^2 = 0.246$, $p<0.0001$; Fig. S11 B). Collectively, this pattern of results offers support for a candidate mechanism in which subnetwork activity tunes functional connectivity. 

\section{Discussion}

\subsection{Functional coupling between default mode and frontoparietal systems tracks individual differences in working memory.}

In a remarkably diverse range of tasks, the frontoparietal system tends to increase in activity while the default mode system tends to decrease in activity \cite{ClareKelly2008,Delaveau2017}. A common and intuitive interpretation of these findings is that the two systems exist in \emph{competition}, consistently displaying anti-correlated dynamics with one another. From a cognitive perspective, it has been proposed that such competition is indicative of opposing goals of the two systems \cite{Boveroux2010,Hannawi2015}, and may enable toggling between the cognitive states supported by each system \cite{Dixon2017}. Importantly, the degree of competition between the frontoparietal and default mode systems has been linked to individual differences in performance on a range of tasks including the categorization of emotion in faces \cite{Xin2014}, the flanker task \cite{ClareKelly2008}, and several tasks taxing human reasoning abilities \cite{Hearne2015}. Interestingly, attenuation of the competition between these two systems is related to decreases in executive function occurring in older age \cite{Spreng2016}.

The nature of the functional connection between the frontoparietal and default mode systems appears to be flexible and task-dependent \cite{Hellyer2014,Elton2014}. Externally-directed tasks tend to elicit competition while internally-directed tasks tend to elicit \emph{cooperation} \cite{Christoff2009,Sridharan2008,Popa2009}. For example, the two systems show competitive dynamics during visuo-spatial planning, and cooperative dynamics during autobiographical planning \cite{Spreng2010}. Likewise, the two systems show cooperative dynamics during a finger-tapping task, and competitive dynamics during a movie watching task \cite{Gao2012}. Similarly, the two systems have been found to show cooperative dynamics during a goal-directed simulation task requiring imagination \cite{Gerlach2011}, as well as during the production of internal trains of thought \cite{Smallwood2012}. It has been proposed that the default mode system is critically involved in self-referential tasks \cite{Andrews-Hanna2014}, and that coupling to the frontoparietal networks allows attention to be directed internally to these tasks, rather than to external stimuli \cite{Smallwood2012,Andrews-Hanna2014}. Our study extends this prior literature by demonstrating the existence of competitive dynamics between the two systems during the performance of an n-back working memory task, where the degree of competition is related to \emph{individual differences} in behavioral performance.

\subsection{Frontoparietal activity is related to its coupling with the default mode system.}

While functional connectivity is estimated from activity time series, the generic relationship between activation and coupling remains far from understood. In the rather non-generic context of working memory tasks, prior studies have shown that the activity of the default mode decreases \cite{Mayer2010,Anticevic2010}, consistent with a suppression of mind-wandering and other self-referential processes that can hamper external goal-directed behavior \cite{Anticevic2012}. In line with these findings, our results demonstrate that during n-back performance, the default mode system is less active on average than the frontoparietal system. Interestingly, we do not observe a relationship between the suppression of default mode activity and individual differences in performance (Fig. S3 B), consistent with a previous study of group-averaged n-back performance in patients with chronic pain and healthy matched controls \cite{Ceko2015}. Indeed, it has been suggested that a threshold may exist, below which working memory performance is unaffected by any further decreases in default mode activity \cite{Ceko2015}. Although our data do not directly support the threshold interpretation, as we observed no significant nonlinearities in the relationship between default mode activity and task accuracy, it would be interesting in future work to further test this hypothesis in groups with markedly lower performance, for example young children or individuals with executive dysfunction.

In contrast to default mode activity, frontoparietal activity directly tracked individual differences in behavior, with higher frontoparietal activation being associated with greater task accuracy (Fig. S3 A). The frontoparietal system is known to support a range of executive functions including trial-by-trial control, enabling the allocation of attention to trial-specific information \cite{Power2013,Cocchi2013}. Such processes are heavily recruited during the n-back task, where successful performance is dependent upon flexibly redirecting attention to similarities and differences between the current stimulus and prior stimuli. More generally, the frontoparietal system is thought to act as a flexible hub during task performance \cite{Cole2013}, altering its connectivity with the default mode system in a task-dependent fashion \cite{Fornito2012,Spreng2010}. Indeed, it has been proposed that the frontoparietal system is functionally (and structurally) interposed between the default mode and dorsal attention systems \cite{Cocchi2013}, modulating their coupling in a task specific way \cite{Dixon2017}. Hellyer and colleagues proposed that increased frontoparietal activity promotes persistent stable states, whereas increased default mode activity may allow for transitions between cognitive states, which in turn would be undesirable during tasks requiring directed and fixed attention \cite{Hellyer2014}. Thus, an active frontoparietal system may also help to maintain a persistent competitive relationship with the default mode, aiding in improved behavioral performance.

Lastly, we investigated the relationship between activity in the frontoparietal and default mode systems and inter-system coupling. It has been hypothesized that competition between the two systems may enable maximally disjunctive computational activities \cite{ClareKelly2008}. While we found that more inter-system competition was related to increased frontoparietal activity, we did not find that functional connectivity between the two systems was related to default mode activity. This pattern of results suggests that competition may not always support maximally disjunctive activities, and instead provides further evidence that -- in the activity range observed here -- default mode activity does not relate to either functional coupling with the frontoparietal system or behavioral performance on the n-back working memory task.

\subsection{Subnetworks of the frontoparietal system modulate coupling to the default mode system.}

The frontoparietal system flexibly alters its functional connections dynamically according to current task demands \cite{Cole2013}, perhaps controlling the strength of connectivity between cognitive systems \cite{Bertolero2018}. In order to support such a broad range of cognitive states, it has been suggested that the frontoparietal system may be composed of subnetworks, where each subnetwork subserves a specific cognitive state \cite{Fornito2012,Dixon2018}. Yet the mechanics of this subservience is far from understood. We proposed that the frontoparietal system is composed of two disjoint subnetworks with anti-correlated dynamics, where subnetwork (A) displays activity that is correlated with the activity of the default mode system, and where subnetwork (B) displays activity that is anti-correlated with the activity of the default mode system. Using an unsupervised clustering algorithm informed by an explicit model of network architecture, we demonstrated that the frontoparietal system is decomposable into two subnetworks with distinct patterns of functional connections. Prior work by Dixon and colleagues has found a similar division of the frontoparietal system \cite{Dixon2018}. The authors employed a hierarchical clustering method to establish a data-driven partition of the frontoparietal system into two components, the first component being more functionally connected to the default mode, similar to our subnetwork (A), and the second component being more functionally connected to the dorsal attention system, similar to our subnetwork (B). Our findings critically extend these prior observations by providing evidence that by supporting competition between the two systems, subnetwork (B) may be critical to working memory performance, while subnetwork (A) may be less involved in working memory, and more closely linked with introspective processes. 

In addition to their functional distinguishability, we also demonstrated that these two subnetworks displayed distinct patterns of gene coexpression. In line with this observation, it is interesting to note that prior work has suggested that cortical regions responsible for different cognitive functions can express different genes \cite{Goldberg2004}, and that gene coexpression provides a partial explanation for patterns of functional connectivity \cite{Richiardi2015}. In agreement with these findings, our results demonstrate more similar patterns of gene expression within subnetworks than between subnetworks, an effect that cannot be explained by inter-regional distance. Notable prior work has also suggested a link between structural connectivity and gene expression \cite{Meyer-Lindenberg2009}, also supported by our finding that the two genetically dissimilar subnetworks display differing patterns of white matter connectivity. It is interesting to speculate that these genetic dissimilarities are partially responsible for the subnetworks' differential capacity to tune the coupling between the frontoparietal and default mode systems via relative changes in activity amplitudes. 

Observing a similar functional division of the frontoparietal system \cite{Fornito2012}, Fornito and colleagues found that increased connectivity between frontoparietal areas and the default mode was correlated with \emph{improved} task performance on an introspective recollection task. Here in a non-introspective task characterized by competitive coupling between the frontoparietal and default mode systems, we observe that increased connectivity between the frontoparietal subnetwork (B) and the default mode system was associated with \emph{poorer} performance. This difference in results can be explained by the observation that the frontoparietal system is a flexible hub \cite{Cole2013}, mediating the connection in particular between the dorsal attention and default mode systems \cite{Cocchi2013}. During introspective tasks it may be advantageous for the frontoparietal system to be more connected to the default mode and less connected to the dorsal attention system, buffering the internal trains of thought from external stimuli \cite{Smallwood2012}. Conversely, during externally-directed tasks (like the n-back), it may be advantageous for the frontoparietal system to be more connected to the dorsal attention system and less connected to the default mode system, buffering externally directed attention from internal thoughts and mind-wandering \cite{Anticevic2012}. In line with this argument, we found that during performance of the 2-back task, subnetwork (B) is positively connected to the dorsal attention system and negatively connected to the default mode system, while the opposite is true for subnetwork (A).

\subsection{Structural connectivity of the frontoparietal subnetworks.}

Motivated by prior work demonstrating that structural and functional connectivity share topographic similarities \cite{Skudlarski2008,Cole2014}, we extend our study to multimodal data to better understand the potential structural drivers constraining the manner in which activity in frontoparietal subnetworks impinges on functional coupling in other systems. We found that subnetwork (A) had fewer structural connections to the dorsal attention system than it had to the default mode system, and also than subnetwork (B) had to the dorsal attention system. Similarly, we found that subnetwork (B) had fewer structural connections to the default mode system than it had to the dorsal attention system, and also than subnetwork (A) had to the default mode system. Collectively, these data suggest that structural connectivity may play a role in constraining the functional dynamics of the two subnetworks. To better understand this role, we draw on tools from network control theory, which provides a mechanistic framework to link network structure to functional network dynamics \cite{Gu2015a,Betzel2016a,Muldoon2016a}. Regions with high levels of boundary control are theoretically posited to have the capacity to steer the brain to different states by coupling and decoupling cognitive systems \cite{Gu2015a}. Our results demonstrate that the frontoparietal subnetworks are situated within the white matter architecture in a manner that can drive the system's coupling with the dorsal attention and default mode systems. The effective activity of this hub of network control may be cognitively advantageous by, during externally-directed tasks, buffering the attentional systems from the internally-directed processes of the default mode, and likewise during internally-directed tasks buffering the default mode system from external stimuli.

\subsection{A reduced dynamical model for probing the relation between system activity and connectivity.}

We proposed a model wherein the coupling between the frontoparietal and default mode systems is modulated by activity levels in the oscillatory dynamics of two frontoparietal subnetworks. To test the validity of this putative mechanism, we employed a simplified dynamical model of brain system activity using oscillators coupled by empirically-determined structural connectivity \cite{Deco2017,Senden2017,Moon2015}. This model allowed us to directly test our hypotheses by allowing us to alter subnetwork amplitude, and subsequently to observe the resulting synchronization (functional connectivity) between the frontoparietal and default mode systems. The results from these simulated experiments agree with our hypotheses and demonstrate that, when reducing the system to a network of 4 coupled oscillators, the modulation of subnetwork amplitude governs intersystem coupling in a way that matches the empirical discoveries. This finding not only provides a mechanistic explanation for our results, but also more generally illuminates how the frontoparietal system may mediate the effective functional coupling between the dorsal attention and default mode systems. 

It has been proposed that the frontoparietal system is functionally interposed between the dorsal attention and default mode systems, altering their functional coupling in a task-specific way. In particular, during externally directed tasks the frontoparietal system may engage with the dorsal attention system and disengage with the default mode system \cite{Smallwood2012}. Conversely, during internally directed tasks the frontoparietal system disengages with the dorsal attention system and engages with the default mode system \cite{Anticevic2012}. Together, these complementary processes are thought to effectively segregate external stimuli from internal trains of thought during tasks that require more focus on one of the two. Our results demonstrate that this complementarity may be accomplished by the differential activation and deactivation of two frontoparietal subnetworks. Indeed, our results suggest that increasing subnetwork (A) activity engages the frontoparietal system with the default mode and increasing subnetwork (B) activity disengages the frontoparietal system from the default mode. Importantly, we found that the opposite relationship is also true when considering the functional coupling between the frontoparietal and dorsal attention systems: increasing subnetwork (A) activity disengages the frontoparietal system with the dorsal attention system and increasing subnetwork (B) activity engages the frontoparietal system with the dorsal attention system. The frontoparietal system is thought to flexibly reconfigure functional connections with the dorsal attention and default mode systems in a task-dependent way \cite{Dixon2017}, and our results offer evidence for a mechanism by which this flexible reconfiguration may be achieved.

\subsection{Methodological Limitations}

Several methodological limitations are pertinent to this work. In the text, the frontoparietal system was divided into two subnetworks, creating group-level subnetworks. First, it should be noted that the subnetworks could also have been studied at the level of single individuals, although such granularity could hamper the ability to draw group-level conclusions. Second, these subnetworks were defined using a community detection algorithm based on a specific generative model. There exist several methods to find communities within networks, each with its own set of underlying assumptions. As a result, careful consideration must be taken when selecting the appropriate method of community detection. The flexibility of the WSBM makes it the most reasonable choice given the data used here. 

While the employed dynamical model indeed recovered empirical results and allowed us to postulate mechanistic explanations, it is important to point out some of its methodological limitations. First, as the most straightforward way to test our main hypotheses, we considered only four major brain systems: the dorsal attention system, default mode system, and the two subnetworks of the frontoparietal system. However, it is crucial to be aware of the fact that these systems are embedded into a larger network, and hence their behavior is determined not only by intrinsic parameters and coupling to one another, but also by interactions with other brain systems. While a comprehensive computational study of large-scale brain dynamics is beyond the scope of this work, it would indeed be interesting in future endeavors to attempt to systematically understand -- via modeling -- the role that other brain systems might have in modulating the activity and functional connectivity of systems related to working memory. 

Another limitation of our model revolves around the scale at which it operates. In particular, because our main empirical findings concern system-level dynamics -- rather than dynamics at the scale of individual neurons or parcels -- we assumed, for simplicity, that the different units in our computational model represented different brain systems. This coarse-grained approach is beneficial for a number of reasons. For example, it simplifies our analysis, and allowed us to focus explicitly on macroscopic, system-level drivers of various results, and therefore directly compare output from the model to corresponding empirical findings. However, although informed by experimental data, it is critical to acknowledge that such a setup is a great simplification of the true system, and allows little room for understanding how observations at the level of brain systems arise from interactions between more microscopic structural components. 

Along this same vein, the dynamics of each brain system were described inherently phenomenologically, via a Hopf bifurcation model \cite{kuznetsov2013elements}, which has been utilized in studies concerned with the interplay between network structure and dynamics in general \cite{Gambuzza2016,Saa2018} and also in computational studies on brain network dynamics more specifically \cite{Deco2017,Senden2017,Moon2015}. In particular, such dynamics indeed capture the oscillatory nature of observed brain system activity, but do not embody a biophysically precise description of neuronal activity. Therefore, the model cannot attempt to describe the emergence of brain system dynamics from dynamical processes on smaller scales. Although in this investigation we have chosen, as a first step, to employ a canonical model with few parameters that favors simplicity and interpretability, building and analyzing more realistic and detailed, multi-scale models is indeed an important area of ongoing research.

\section{Conclusion}

Our study provides evidence for a mechanism by which the dynamics of the frontoparietal system may drive working memory performance. Two distinct subnetworks within the frontoparietal system play a role in modulating the functional coupling between the frontoparietal and default mode systems during the performance of an n-back working memory task, which may help buffer the externally-directed attentional system from internal trains of thought, and lead to improved behavioral performance. We found that the position of the two subnetworks within the white matter scaffolding constrains the distinct function of each: one is structurally tied to the dorsal attention system (and drives the frontoparietal system into competition with the default mode), and the other is structurally tied to the default mode (and drives the frontoparietal system into cooperation with the default mode). We extend these descriptive observations by building a computational model instantiating and validating the putative mechanism, and we bolster our findings with corroborating differences in gene expression in the two subnetworks. Together, our findings contribute to a holistic view of working memory by linking activity, functional connectivity, structural connectivity, and gene expression, and present one way of understanding how these four modes work in concert to support cognitive processes necessary for working memory.

\section*{Acknowledgments}

We thank Karolina Finc, Lorenzo Caciagli, and Ursula Tooley for helpful comments on an earlier version of this manuscript. Our team acknowledges support from the John D. and Catherine T. MacArthur Foundation, the Alfred P.  Sloan  Foundation,  the  ISI  Foundation,  the  Paul  G.  Allen  Foundation,  the  Army  Research  Laboratory  (W911NF-10-2-0022),  the  Army  Research  Office  (Bassett-W911NF-14-1-0679,  Grafton-W911NF-16-1-0474, DCIST- W911NF-17-2-0181), the Office of Naval Research, the National Institute of MentalHealth (2-R01-DC009209-11, R0-MH112847, R01-MH107235, R21-MH106799, R01-MH113550), the National  Institute  of  Child  Health  and  Human  Development  (1R01-HD086888-01),  National  Institute  of Neurological Disorders and Stroke (R01-NS099348), and the National Science Foundation (BCS-1441502,BCS-1430087,  NSF  PHY-1554488  and  BCS-1631550).  The  content  is  solely  the  responsibility  of  the authors and does not necessarily represent the official views of any of the funding agencies.

\clearpage
\newpage
\bibliographystyle{plain}

\end{document}


\doublespacing
\noindent \large{\textbf{Supplement to:\\ ``Multiscale and multimodal network dynamics underpinning working memory''}}
\\
\\
Andrew C. Murphy\textsuperscript{1,2}, Maxwell A. Bertolero\textsuperscript{1}, Lia Papadopoulos\textsuperscript{3}, David M. Lydon-Staley\textsuperscript{1}, and Danielle S. Bassett\textsuperscript{1,3,4,5,6,*}
\\
\\
\textsuperscript{1}Department of Bioengineering, School of Engineering \& Applied Science, University of Pennsylvania, Philadelphia, PA 19104, USA
\\
\textsuperscript{2}Perelman School of Medicine, University of Pennsylvania, Philadelphia, PA 19104, USA
\\
\textsuperscript{3}Department of Physics \& Astronomy, School of Arts \& Sciences, University of Pennsylvania, Philadelphia, PA 19104, USA
\\
\textsuperscript{4}Department of Neurology, Perelman School of Medicine, University of Pennsylvania, Philadelphia, PA 19104, USA
\\
\textsuperscript{5}Department of Electrical \& Systems Engineering, School of Engineering \& Applied Science, University of Pennsylvania, Philadelphia, PA 19104, USA
\\
\textsuperscript{6}Department of Psychiatry, Perelman School of Medicine, University of Pennsylvania, Philadelphia, PA 19104, USA
\\
\textsuperscript{*}To whom correspondence should be addressed: dsb@seas.upenn.edu

\clearpage
\setcounter{figure}{0}   
\makeatletter 
\renewcommand{\thefigure}{S\@arabic\c@figure}
\makeatother 
\newpage
\begin{figure}[H]
\centerline{\includegraphics{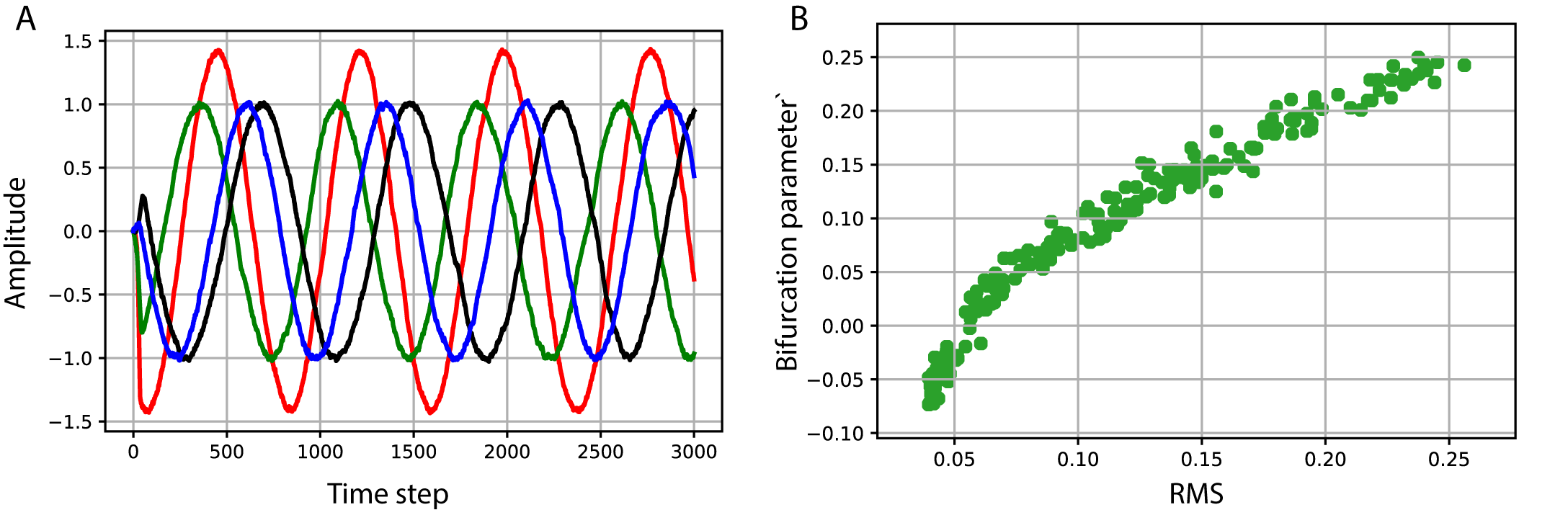}}
\caption{\textbf{Relationship between signal amplitude and bifurcation parameter from the computational model of brain system dynamics.} \emph{(A)} Here we show a simulation of a 4 node network. The dynamics of each unit are described by the normal form of a Hopf bifurcation (see Methods in the main text), and exhibit oscillatory activity at the chosen bifurcation parameters. The green, blue, and black oscillators have identical bifurcation parameters equal to 1. The red oscillator has a bifurcation parameter equal to 2, which serves to increase the amplitude without changing other features of the signal. \emph{(B)} Here we simulate a 4 node network while varying the bifurcation parameter of one node, and measuring the root mean square (RMS) of the resulting signal of the same node to show that the manipulation of the bifurcation parameter linearly alters the RMS of the resulting time series (Pearson correlation coefficient $r^2 = 0.93$, $p < 0.0001$).}\label{figS9}
\end{figure}

\begin{figure}[H]
\centerline{\includegraphics{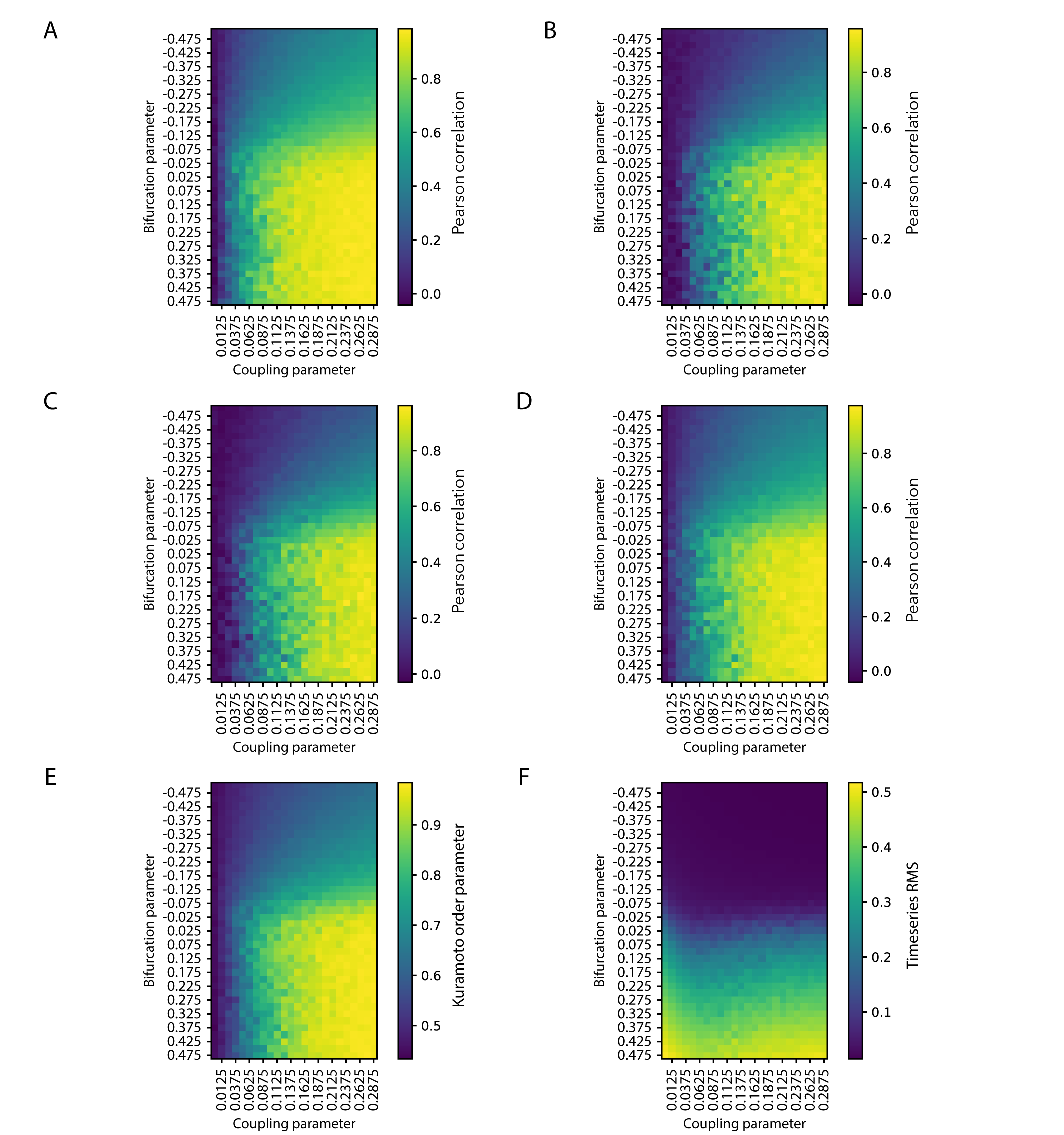}}
\caption{\textbf{Determination of Hopf bifurcation model parameters.} \emph{(A)} Strength of functional connectivity between subnetwork (A) and the default mode system as a function of the coupling parameter and of the bifurcation parameter. \emph{(B)} Strength of functional connectivity between subnetwork (A) and the dorsal attention system as a function of the coupling parameter and of the bifurcation parameter. \emph{(C)} Strength of functional connectivity between subnetwork (B) and the default mode system as a function of the coupling parameter and of the bifurcation parameter. \emph{(D)} Strength of functional connectivity between subnetwork (B) and the dorsal attention system as a function of the coupling parameter and of the bifurcation parameter. \emph{(E)} Kuramoto order parameter as a function of the coupling parameter and of the bifurcation parameter. \emph{(F)} Mean time series RMS as a function of the coupling parameter and of the bifurcation parameter.}\label{figS10}
\end{figure}

\begin{figure}
\centerline{\includegraphics[]{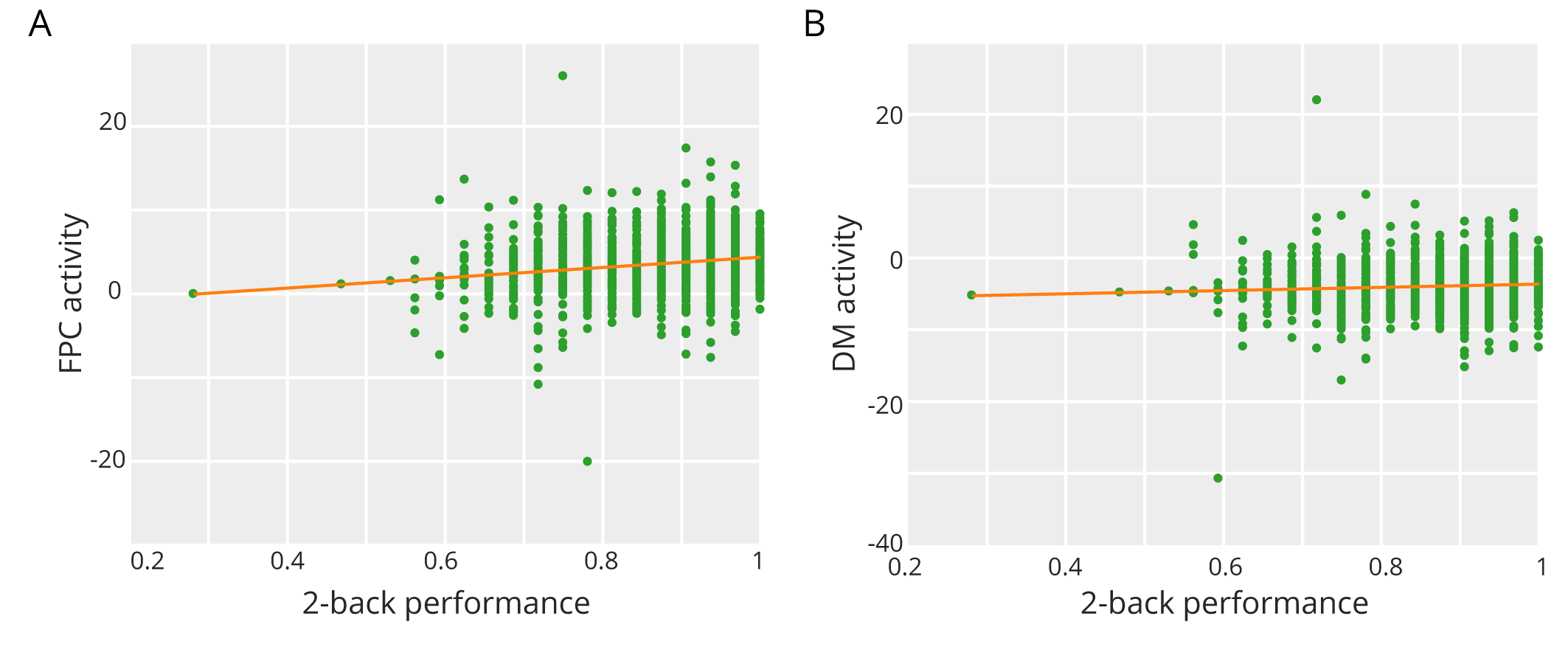}}
\caption{\textbf{Frontoparietal activity correlates with working memory performance.} During the 2-back working memory task, we found that the activity of the frontoparietal system is related to behavioral performance (see panel \emph{(A)}; repeated measures correlation; $r =  0.1019$, $p = 0.016$, $DF = 558$), while the activity of the default mode system is unrelated to behavioral performance (see panel \emph{(B)}; repeated measures correlation; $r =  0.0384$, $p = 0.363$, $DF = 558$).} \label{figS2}
\end{figure}

\section*{Supplementary note 1: Alternate behavioral measure}
\begin{figure}[h]
\centerline{\includegraphics{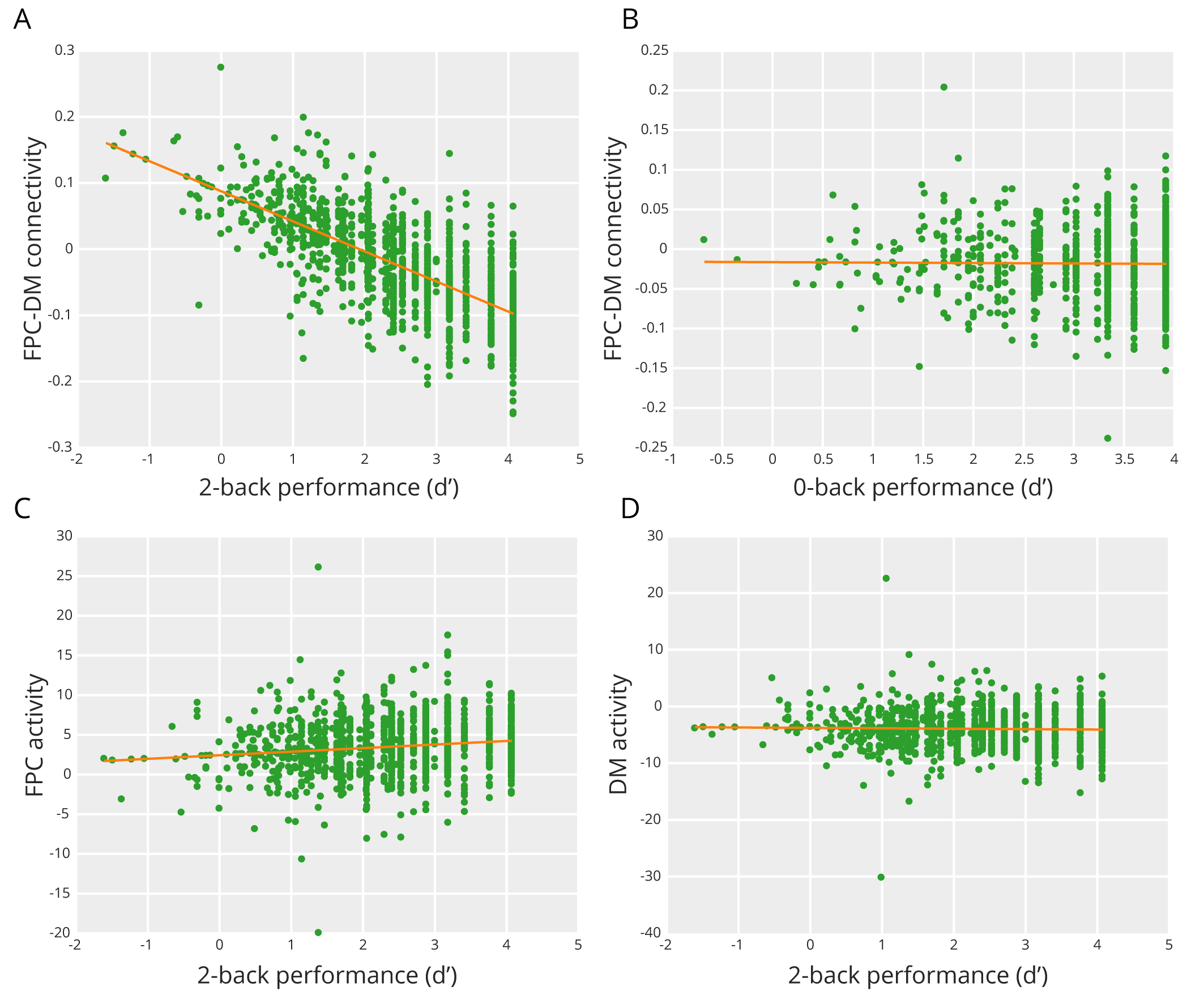}}
\caption{\textbf{The relationship of connectivity and activity to behavior (d').} Using the alternate behavioral metric of d', the relationship of behavior to both connectivity and activity remained unchanged. Specifically, the functional coupling between the frontoparietal and default mode systems during the 2-back task is negatively correlated with performance (see panel \emph{(A)}; repeated measures correlation; $r = -0.4738$, $p < 0.001$, $DF = 562$), and the functional coupling during the 0-back task is not significantly correlated with performance (see panel \emph{(B)}; repeated measures correlation; $r =  -0.0195$, $p = 0.643$, $DF = 562$). Similarly, during the 2-back task the frontoparietal activity is marginally correlated with performance (see panel \emph{(C)}; repeated measures correlation; $r =  0.0828$, $p =0.049$, $DF = 562$), while the default mode activity is not significantly correlated with performance (see panel \emph{(D)}; repeated measures correlation; $r = -0.0139$, $p = 0.74$, $DF = 562$).}\label{figS5}
\end{figure}

In the main text, behavioral performance on the n-back working memory task was defined as the total accuracy, measured by the ratio of correct responses to total responses. To probe the robustness of our results, we examined an alternate behavioral metric given by the sensitivity index (d'). The sensitivity index quantifies a subject's ability to discriminate between previously seen and novel stimuli \cite{Snodgrass1988} and is sometimes used as an alternative to accuracy in the study of behavioral performance on working memory tasks \cite{Satterthwaite2013}. The d' metric and accuracy are correlated with one another during the 0-back task (repeated measures correlation; $r = 0.89$, $p < 0.001$, $DF = 562$) and during the 2-back task (repeated measures correlation; $r = 0.852$, $p < 0.001$, $DF = 558$). Using this alternate metric, we found that the relationship between behavior and the functional coupling between the frontoparietal and default mode systems remained unchanged. Furthermore, using this alternate metric, we found that the relationship between behavior and the activity of the frontoparietal system remained unchanged, as did the relationship between behavior and the activity of the default mode system (Fig. \ref{figS5}).

\clearpage
\section*{Supplementary note 2: Alternate parcellation}
In the main text, we chose to divide the brain into 400 discrete non-overlapping regions where each region is assigned to a functional system. However, recent studies commonly report results across at least two parcellations, as it remains unclear whether a single parcellation is optimal for all studies and the testing of all hypotheses. While there have been several attempts to empirically define the optimal number of brain areas \cite{Eickhoff2015,Yeo2011}, the brain is organized hierarchically \cite{Blumensath2013}, and so the number and size of parcellations to use becomes dependent on the question to be addressed. One of the requirements we imposed on our choice of parcellation was the ability to discern inhomogeneity in the frontoparietal system, and so we required subdivisions within that system. To reduce parcellation complexity, we also wanted a small number of total brain parcels. The 400 region parcellation provided us with a relatively small total number of parcels, while maintaining a relatively high resolution within the frontoparietal system (61 parcels). However, we were interested to see how robust our results were with respect to the choice of parcellation, and thus we considered a second, lower resolution parcellation composed of 100 regions \cite{Schaefer2017}. Because of the difference in resolution, we expected that system-level results would be consistent across parcellations, but that subnetwork results might be altered.

\begin{figure}[H]
\centerline{\includegraphics{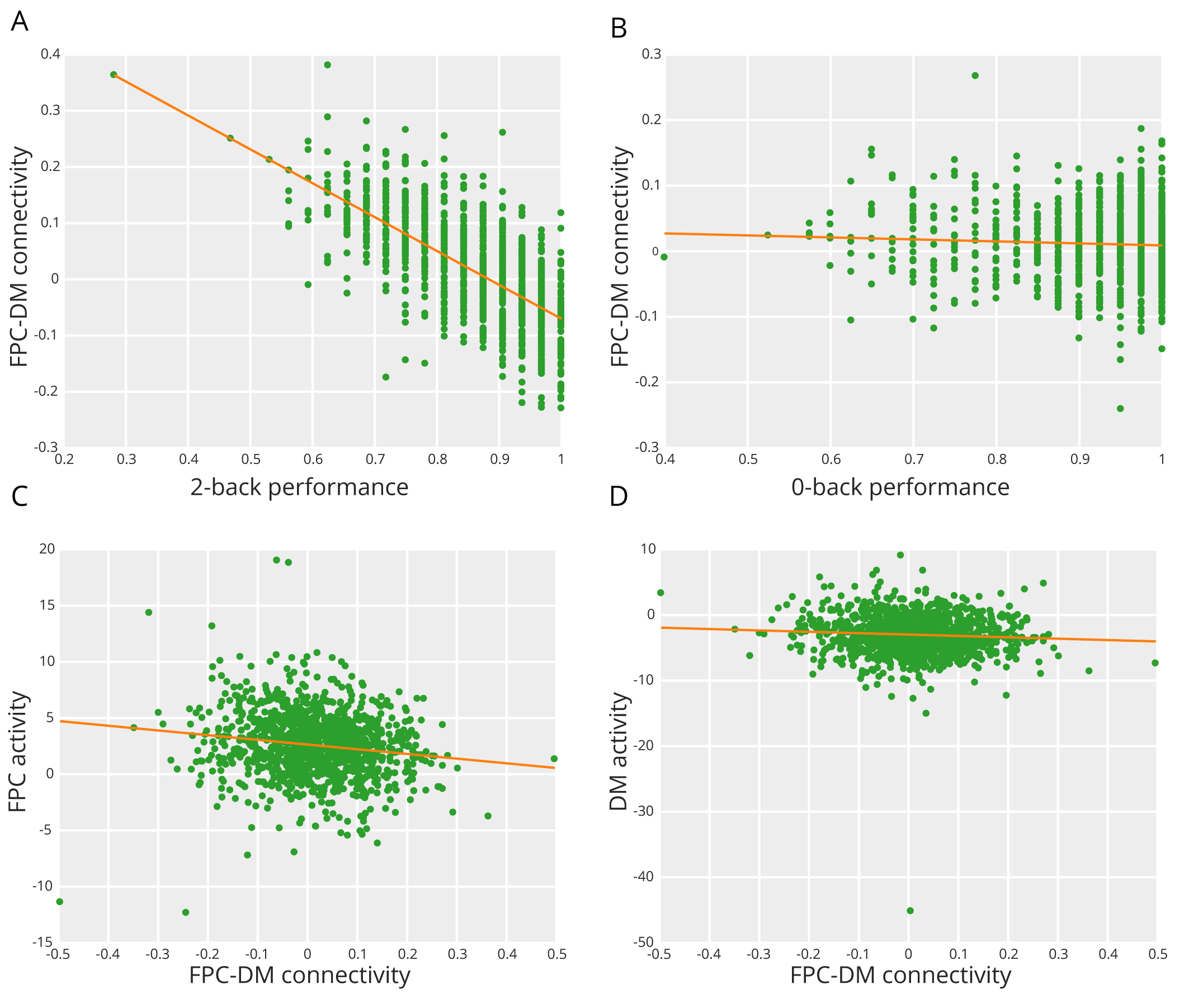}}
\caption{\textbf{Replication of results with a different parcellation.} Using the alternate parcellation, the relationships between connectivity and behavior remained unchanged. Specifically, the functional coupling between the frontoparietal and default mode systems during the 2-back working memory task was negatively correlated with performance (see panel \emph{(A)}; repeated measures correlation; $r = -0.4498$, $p < 0.001$, $DF = 558$), and the functional coupling between the two systems during the 0-back task was not significantly correlated with performance (see panel \emph{(B)}; repeated measures correlation; $r = -0.0347$, $p = 0.409$, $DF = 562$). Similarly, during the 2-back task the frontoparietal activity was negatively correlated with performance (see panel \emph{(C)}; repeated measures correlation; $r = -0.0835$, $p = 0.0473$,$DF = 562$), while the default mode activity was not significantly correlated with performance (see panel \emph{(D)}; repeated measures correlation; $r = -0.0745$, $p = 0.0769$, $DF = 562$).}\label{figS6}
\end{figure}

\begin{figure}[H]
\centerline{\includegraphics[scale=0.8]{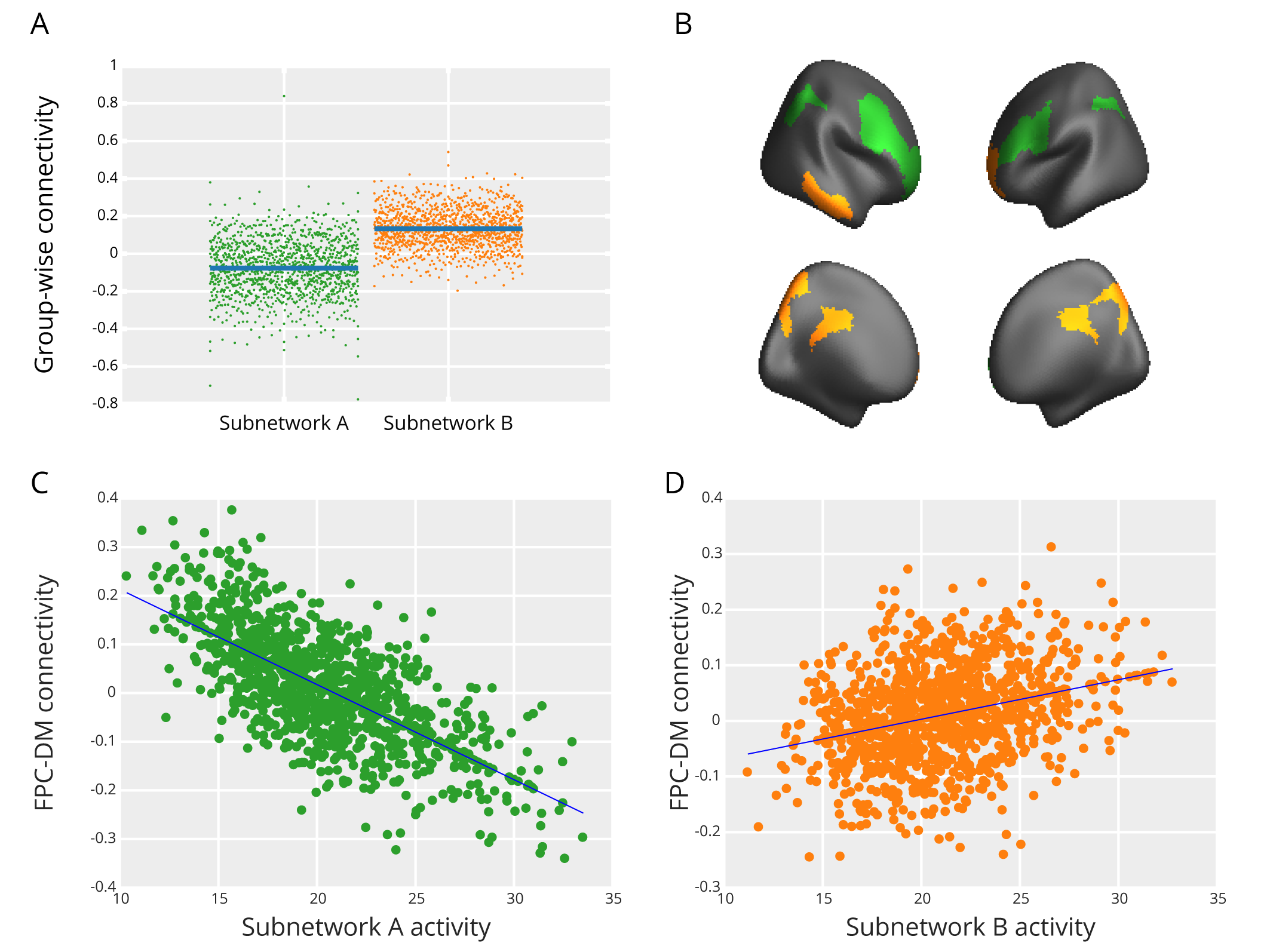}}
\caption{\textbf{Replication of subnetwork results with a different parcellation.} \emph{(A)} Using the alternate parcellation, we found that two frontoparietal subnetworks exist with different relationships to the default mode system. Specifically, we found that subnetwork (A) displays activity that is negatively correlated with the activity of the default mode system (mean $r= -0.076$, 95\% CI: (-0.084,-0.068)), while subnetwork (B) displays activity that is positively correlated with the activity of the default mode system (mean $r= 0.132$, 95\% CI: (0.126, 0.138)). The two groups significantly differ from each other according to both our multilevel model ($\beta = -0.20892$, $p<0.0001$, $t(1769) = -41.3$, $SE = 0.0050548$, $n = 2414$), and according to a non-parametric permutation test ($p<0.0001$). \emph{(B)} The two groups projected onto the cortical surface. \emph{(C)} Subnetwork (A) activity is negatively correlated with the functional coupling between the frontoparietal and default mode systems ($\beta = -0.0088$, 95\% CI: (-0.00989, -0.00787)). \emph{(D)} Subnetwork (B) activity is positively correlated with the functional coupling between the frontoparietal and default mode systems ($\beta = 0.00435$, 95\% CI: (0.00298, 0.00571)) when using both as independent variables in a robust linear mixed effects model.
}\label{figS7}
\end{figure}

In line with our hypothesis, we found that the system-level results were unaltered (Fig.~\ref{figS6}), while the subnetwork level results did change modestly as a function of parcellation. The frontoparietal system can still be separated into two distinct subnetworks where one displays activity that is correlated with the activity of the default mode system (subnetwork (B)) and the other displays activity that is anticorrelated with the activity of the default mode system (subnetwork (A)) (Fig.~\ref{figS7}A, B). Furthermore, as before, increasing subnetwork (A) activity is related to decoupling of the frontoparietal and default mode systems (Fig.~\ref{figS7}C), and increasing subnetwork (B) activity is related to coupling of the frontoparietal and default mode systems (Fig.~\ref{figS7}D). The primary difference between results obtained from the 400 region parcellation and from the 100 region parcellation is the anatomical distribution of the subnetworks over the cortical surface. Whereas before we observed subnetworks that were partially constrained to different hemispheres, we now observe sub-networks constrained to either the medial (subnetwork (B)) or lateral (subnetwork (A)) surfaces. These findings suggest that the large-scale architecture of the 100 region parcellation is more sensitive to differences in dynamics that are maintained across both hemispheres, while the smaller-scale architecture of the 400 region parcellation is more sensitive to differences in dynamics that vary across the two hemispheres. More generally, it is well known that including a smaller number of regions, where regions are larger on average, enables the examination of larger hierarchical dynamics \cite{Schaefer2017}, resulting in the identification of different subnetworks. 

It is important to emphasize that this division of the frontoparietal system is no less valid than the one presented in the main text, but rather is reflective of dynamics on a larger physical scale. Indeed, it has been reported that the medial parietal cortex activates during the observance of social interactions \cite{Iacoboni2004}, which requires recruitment of the default mode system \cite{Mars2012}. The medial parietal cortex is part of our subnetwork (B), which is heavily integrated with the default mode system and drives functional coupling between the frontoparietal and default mode systems. Furthermore our subnetwork (A) is largely composed of the prefrontal cortex, a region long established as critical to working memory \cite{Kane2002,Curtis2003,Balconi2013}, and drives the anticorrelations between the frontoparietal and default mode systems, which in turn could provide the necessary functional buffer from internal trains of thought.

\section*{Supplementary note 3: Alternative functional connectivity measure}
\begin{figure}[H]
\centerline{\includegraphics[width=6.8in]{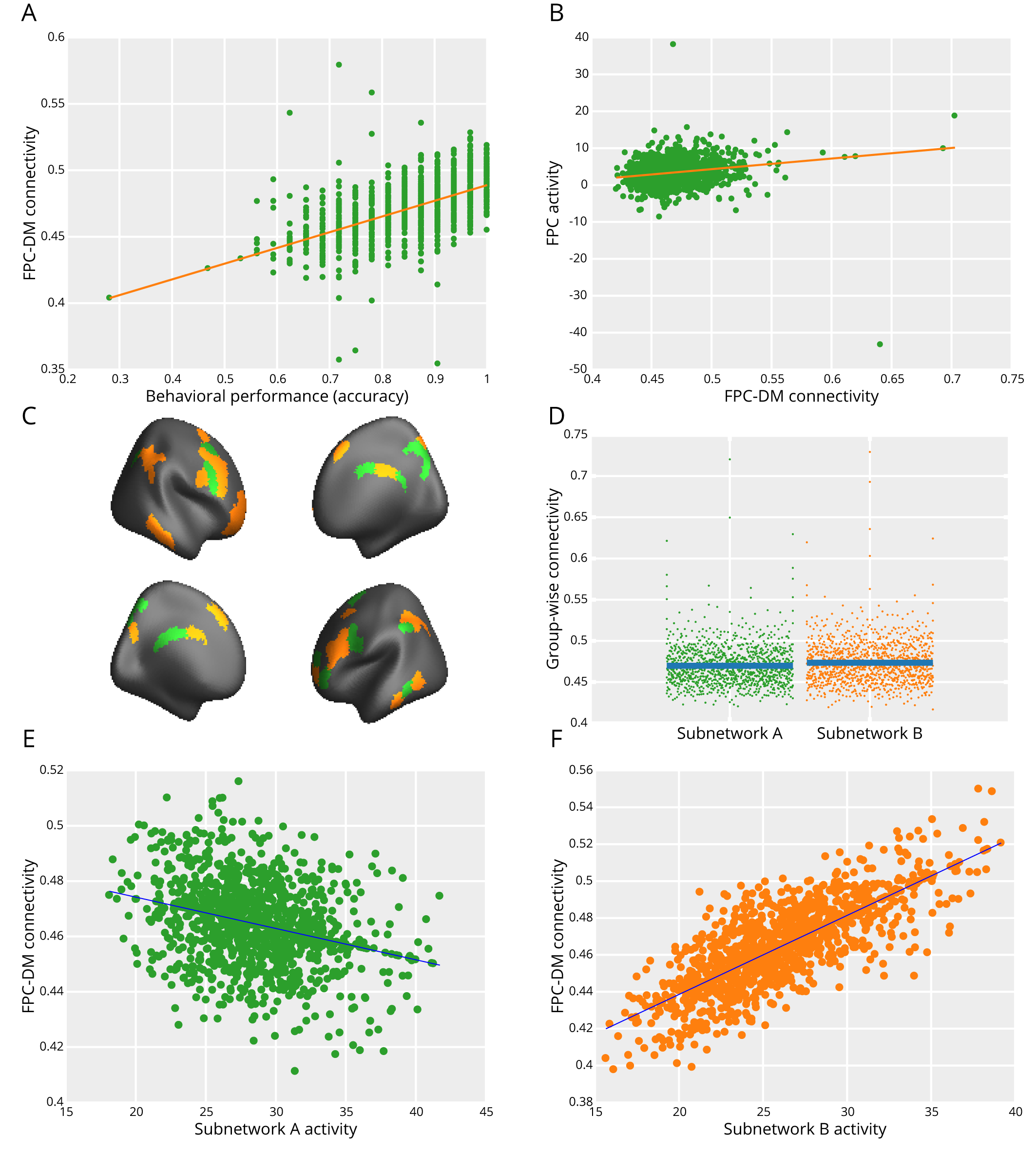}}
\caption{\textbf{Replication of results with a coherence-based estimate of functional connectivity.} \emph{(A)} Behavioral performance on the 2-back working memory task relates to the functional coupling between the frontoparietal and default mode systems (repeated measures correlation; $r = 0.356$, $p <  0.0001$, $DF = 555$). \emph{(B)} Frontoparietal activity relates to the functional coupling between the frontoparietal and default mode systems (repeated measures correlation; $r = 0.0968$, $p = 0.0299$, $DF = 500$). \emph{(C)} Community detection via the weighted stochastic block model reveals a 2 subnetwork fractioning of the frontoparietal system. \emph{(D}) Subnetwork (A) is less connected to the default mode system (mean = 0.469, 95\% CI: (0.468, 0.471)) than subnetwork (B) (mean = 0.473, 90\% CI: (0.472, 0,475)). The two groups significantly differ from each other according to both a multilevel model ($\beta = 0.0038179$, $p<0.0001$, $t(1769) = 4.2$, $SE = 0.00089474$, $n = 2414$), and a non-parametric permutation test ($p<0.0001$). Fitting a single robust model accounting for the activity of both subnetworks demonstrates \emph{(E)} that subnetwork (A) activity is negatively correlated with the functional coupling between the frontoparietal and default mode systems ($\beta = -0.003055$, 95\% CI: (-0.00056, -0.000005)), while \emph{(F)} subnetwork (B) activity is positively correlated with the functional coupling between the frontoparietal and default mode systems ($\beta = 0.00148$, 95\% CI: (0.001246, 0.000172)).}\label{figS8}
\end{figure}

In the main text, we computed functional connectivity using the Pearson correlation coefficient. However, there are numerous methods used to establish functional connectivity between time series in neuroimaging \cite{Wang2014}, with correlation-based measures accounting for just a single subset of these methods. In order to demonstrate the robustness of our results, we sought to reproduce our main findings using a measure of functional connectivity unrelated to the correlation. We chose to use wavelet coherence due to its established usage in neuroimaging \cite{Sun2004,Bassett2013,Bassett2015,Murphy2016}, and we chose wavelet scale 2, which corresponds to the relevant frequency range of 0.06 Hz - 0.12 Hz \cite{Bassett2013}. Rather than probing a single linear relationship between time series, coherence is a measure of the cross-correlation between time series \cite{Wang2014}. Similar to the Pearson correlation coefficient, the wavelet coherence is undirected. However, unlike the Pearson correlation coefficient, wavelet coherence is bounded in [0,1], and thus anti-correlation cannot be observed. Importantly, our main findings hold using this different estimate of functional connectivity (Fig. \ref{figS8}).

Specifically, we found that the strength of the functional connection between the default mode and frontoparietal systems remains related to behavioral performance on the 2-back working memory task ($\beta = 0.11817$, $p <  0.0001$, $t(555) = 6.2466$, $SE = 0.018844$, $n = 1192$; Fig.~\ref{figS8}A), and that the strength of the intersystem connection remains correlated with frontoparietal activity ($\beta =  28.759$, $p = 0.0002$, $t(560) = 3.7279$, $SE = 7.7145$, $n = 1203$; Fig.~\ref{figS8}B). The sign of the relationship between the functional connection strength and behavioral performance is inverted relative to the Pearson correlation results, due to the fact that coherence is always positive. Community detection to identify two distinct subnetworks of the frontoparietal system resulted in a slightly different split than that reported in the main manuscript (Fig.~\ref{figS8}C), although the two share some general characteristics. Importantly, the coherence-derived subnetworks are functionally distinct, with subnetwork (A) less strongly connected to the default mode than subnetwork (B). Furthermore, in keeping with our main results, (i) the activity of the subnetwork more weakly connected to default mode (subnetwork (A)) displays activity that is negatively correlated with the functional coupling between the frontoparietal and default mode systems (Fig.~\ref{figS8}E), and (ii) the activity of the subnetwork more strongly connected to the default mode system (subnetwork (B)) displays activity that is positively correlated with the functional coupling between the default mode and frontoparietal systems (Fig.~\ref{figS8}F).

\clearpage
\section*{Supplementary note 4: Accounting for the effect of spatial distance on gene coexpression between brain regions}
\label{gene_distances}
Gene coexpression is known to decrease as a function of distance between the two regions being studied \cite{pantazatos2017commentary}. Thus, any observed difference in gene coexpression between two subnetworks could be due rather trivially to differences in distance spanned by the regions in the two subnetworks. To ensure that distance was not driving our observations, we assessed the pairwise distances between regions in subnetwork (A), the pairwise distances between regions in subnetwork (B), and the pairwise distances between all regions in the frontoparietal system, which is comprised of regions in both subnetwork (A) and subnetwork (B). We defined the pairwise distance between two parcels to be the Euclidian distance between the mean X, Y, and Z coordinates of the respective parcels in volumetric space. Specifically, we drew 50000 bootstrap samples of regions in subnetwork (A), subnetwork (B), and the frontoparietal system as a whole. We then calculated (i) the difference between the mean subnetwork (A) distance and the mean subnetwork (B) distance, (ii) the difference between the mean subnetwork (A) distance and the mean frontoparietal system distance, and (iii) the difference between the mean subnetwork (B) distance and the mean frontoparietal system distance. To conclude the analysis, we determined whether the 95\% confidence interval for any of these difference distributions included zero.

After performing these computations, we found that the difference in mean within-subnetwork distance between subnetwork (A) and subnetwork (B), as well as the difference in mean within-subnetwork distance between subnetwork(A) (or (B)) and the mean frontoparietal network distance, were all not significantly different than 0, after Bonferroni correction for multiple comparisons. Specifically, we found that the difference in mean within-subnetwork distance between subnetwork (A) and subnetwork (B) is not significantly different than 0 $(95\%~CI: [ -8.5979, 3.7972])$, the difference in mean within-subnetwork distance between subnetwork (A) and mean within frontoparietal network distance is not significantly different than 0 $(95\%~CI: [-6.6320, 2.5436])$, and the difference in mean within-subnetwork distance between subnetwork (B) and the mean frontoparietal network distance is not significantly different than 0 $(95\%~CI: [-9.6414, 0.8261])$. Given these findings, inter-regional distance is unlikely to explain the observed differences in gene coexpression in the two subnetworks.

\begin{figure}
\centerline{\includegraphics[]{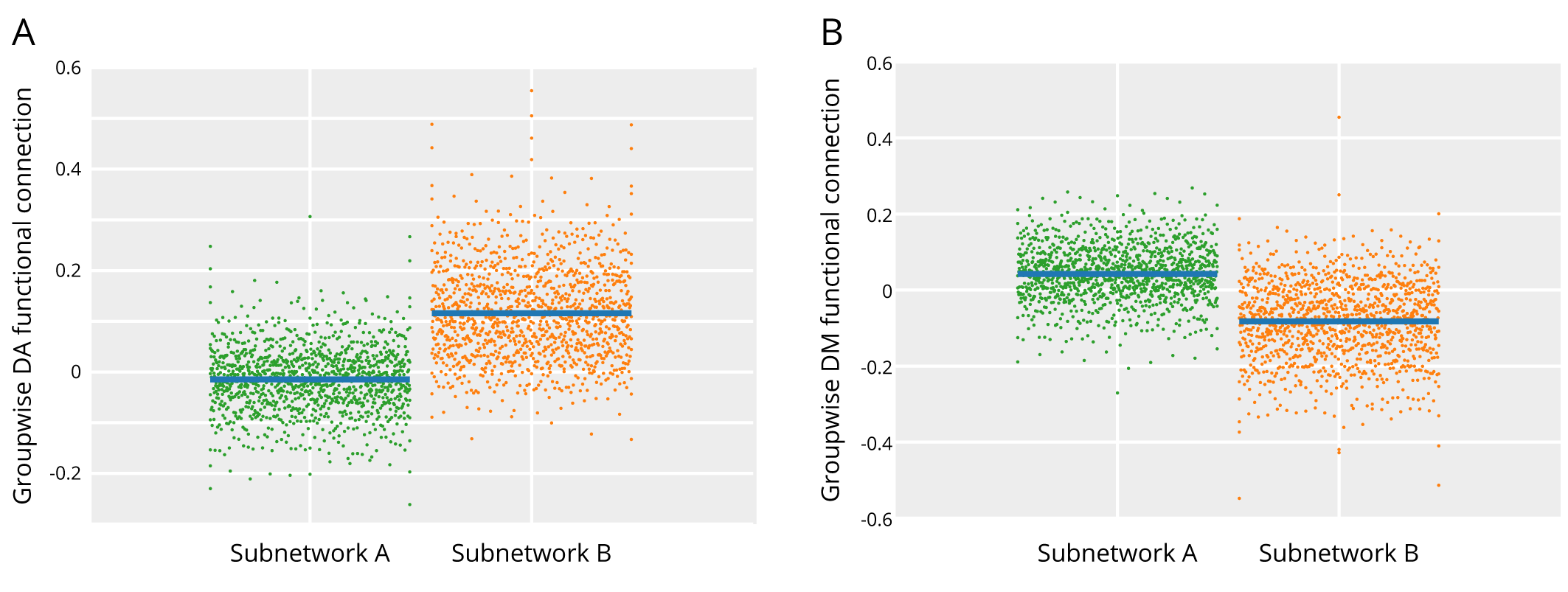}}
\caption{\textbf{Frontoparietal subnetwork connectivity.} \emph{(A)} Subnetwork (A) displays activity that is negatively correlated with the activity of the dorsal attention system (mean$r = -0.14$, 95\%CI: (-0.18, -0.11)), while subnetwork (B) displays activity that is positively correlated with the activity of the dorsal attention system (mean $r = 0.12$, 95\%CI: (0.11, 0.12)). Using a multilevel model, we found that these correlations are significantly different ($\beta = 0.13057$, $p<0.0001$, $t(1769) = 46.8$, $SE =  0.0027885$, $n = 2414$), and we found similar results using a non-parametric permutation test ($p<0.0001$). \emph{(B)} Subnetwork (A) displays activity that is positively correlated with the activity of the default mode system (mean $r= 0.042$, 95\%CI: (0.038, 0.046)), while subnetwork (B) displays activity that is negatively correlated with the activity of the default mode system (mean $r= -0.082$, 95\%CI: (-0.088, -0.076)). Using a multilevel model, we found that these correlations are significantly different ($\beta = -0.12469$, $p<0.0001$, $t(1769) = -37.4$, $SE = 0.0033272$, $n = 2414$), and we found similar results using a non-parametric permutation test ($p<0.0001$). Note that for visualization purposes, the data was not visually adjusted to account for subject repeated measures effects, whereas all reported statistics do take this into account.}\label{figS3}
\end{figure}
\newpage
\clearpage
\begin{figure}
\centerline{\includegraphics[]{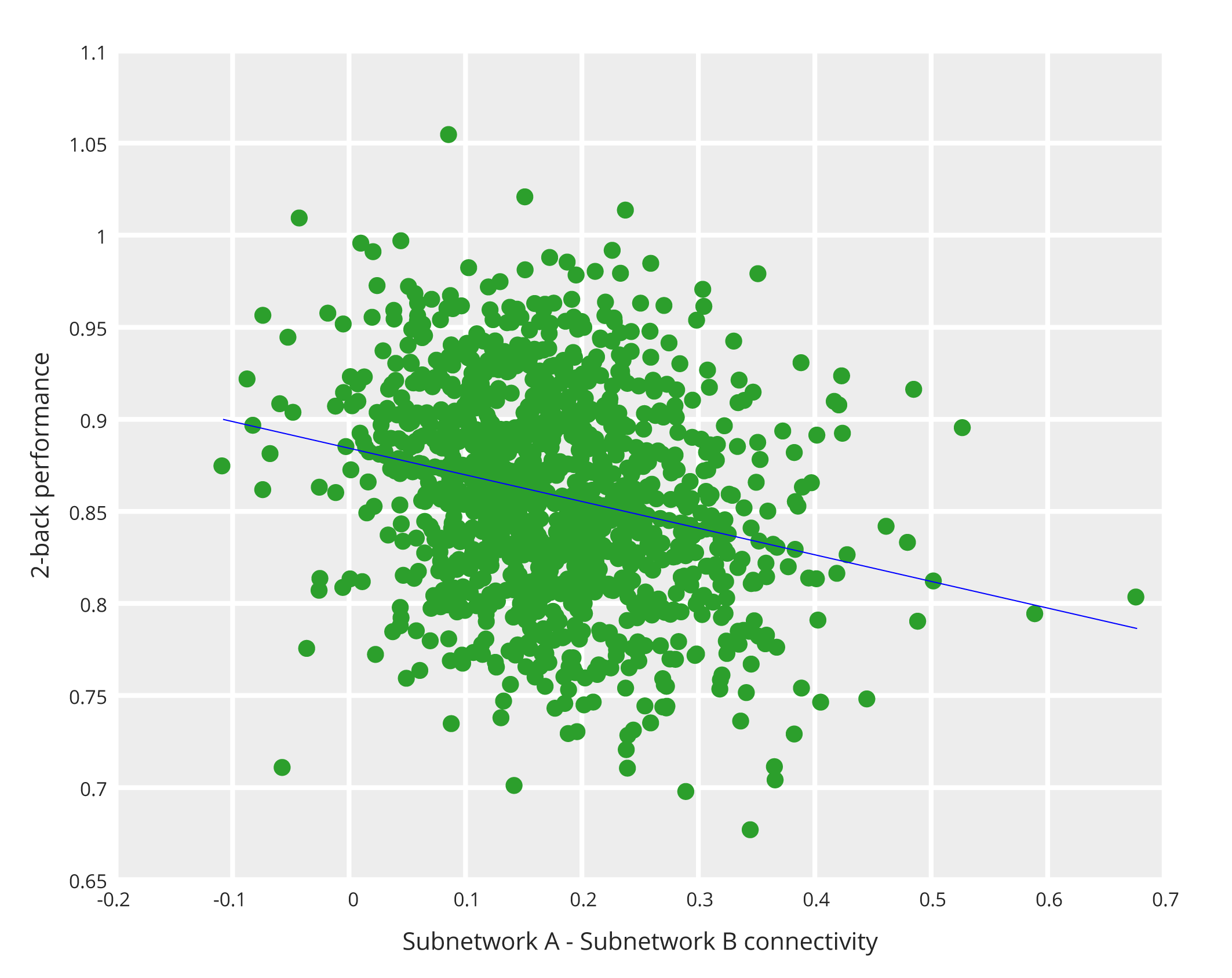}}
\caption{\textbf{Frontoparietal subnetwork connectivity is related to performance.} As the two frontoparietal subnetworks display more correlated time series, performance on the 2-back working memory task worsens (Pearson correlation coefficient $r^2 = 0.73$, $p = 0.0003$).}\label{figS4}
\end{figure}

\begin{figure}[H]
\centerline{\includegraphics{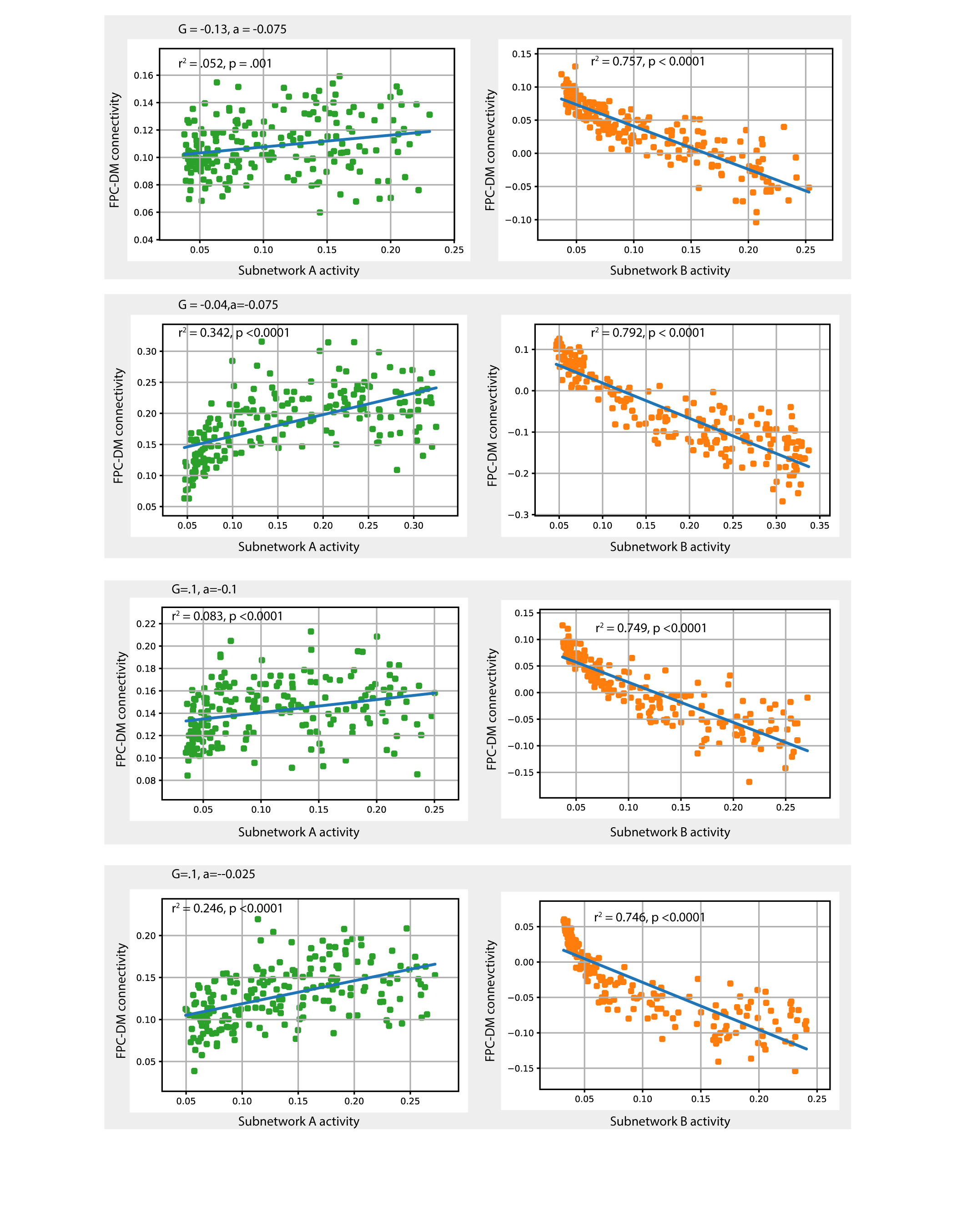}}
\caption{\textbf{Results from the dynamical model at different parameter values.} These figures reproduce the results of Fig. 5 in the main manuscript at different values of the coupling and bifurcation parameters for the 3 nodes where activity is not being modulated.}\label{figS12}
\end{figure}

\begin{figure}[H]
\centerline{\includegraphics{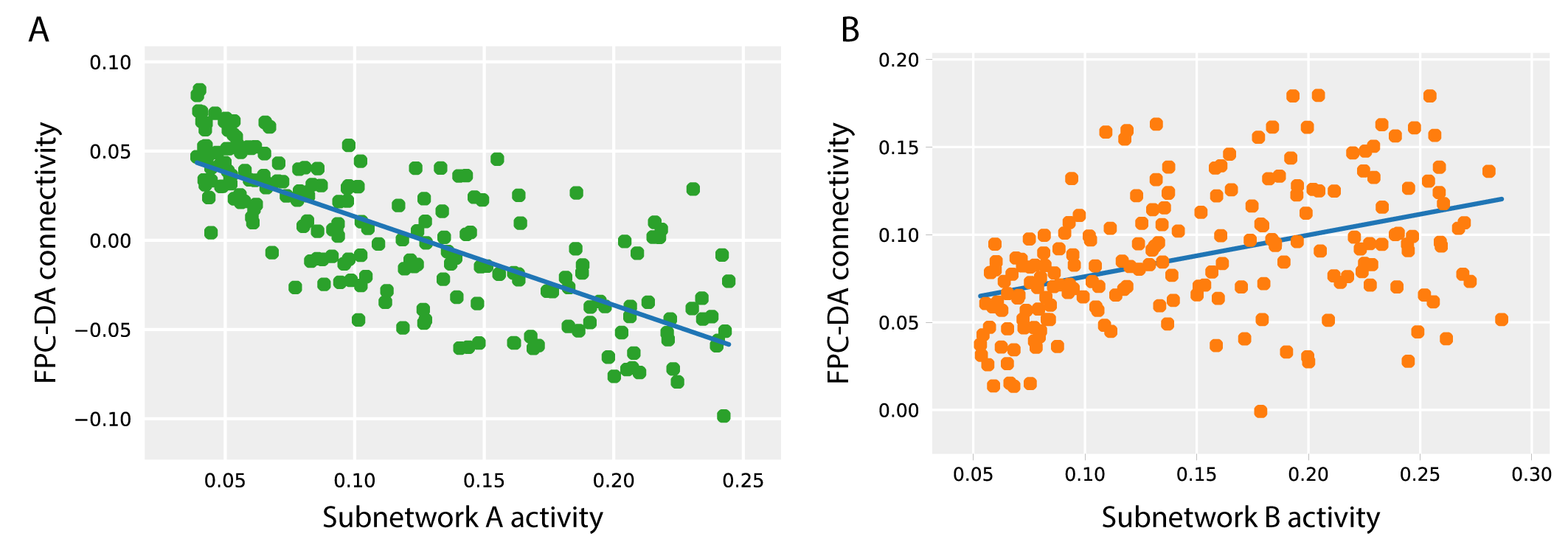}}
\caption{\textbf{Simulated frontoparietal and dorsal attention dynamics.} \emph{(A)} Increasing the amplitude of subnetwork (A) in the computational model caused a decrease in the functional coupling between the frontoparietal and dorsal attention systems (Pearson correlation coefficient $r^2 = 0.65$, $p < 0.0001$). \emph{(B)} Increasing the amplitude of subnetwork (B) in the computational model caused an increase in functional coupling between the frontoparietal and dorsal attention systems (Pearson correlation coefficient $r^2 = 0.246$, $p < 0.0001$).}\label{figS11}
\end{figure}

\begin{figure}[H]
\centerline{\includegraphics{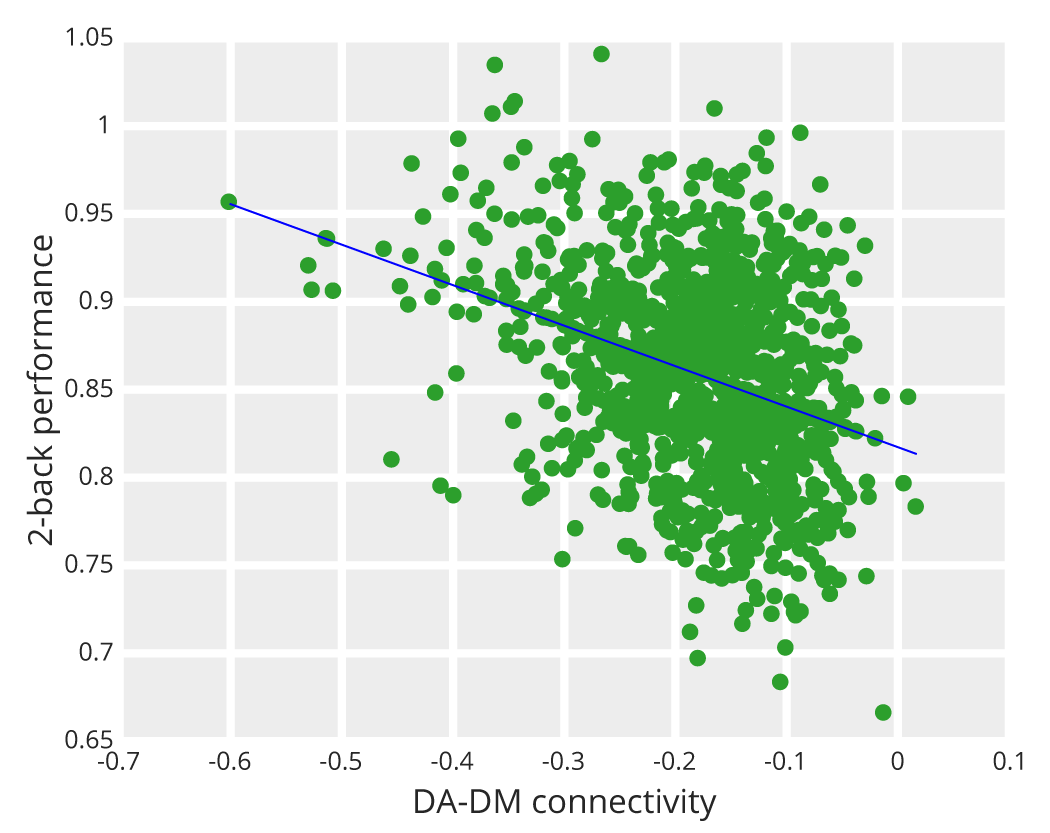}}
\caption{\textbf{The strength of functional coupling between the dorsal attention and default mode systems relates to behavior.} The strength of the functional connection between the default mode and dorsal attention systems is anti-correlated with behavioral performance (Pearson's correlation coefficient $r^2 = 0.73$, $p < 0.0001$).}\label{figS13}
\end{figure}

\begin{figure}[H]
\centerline{\includegraphics{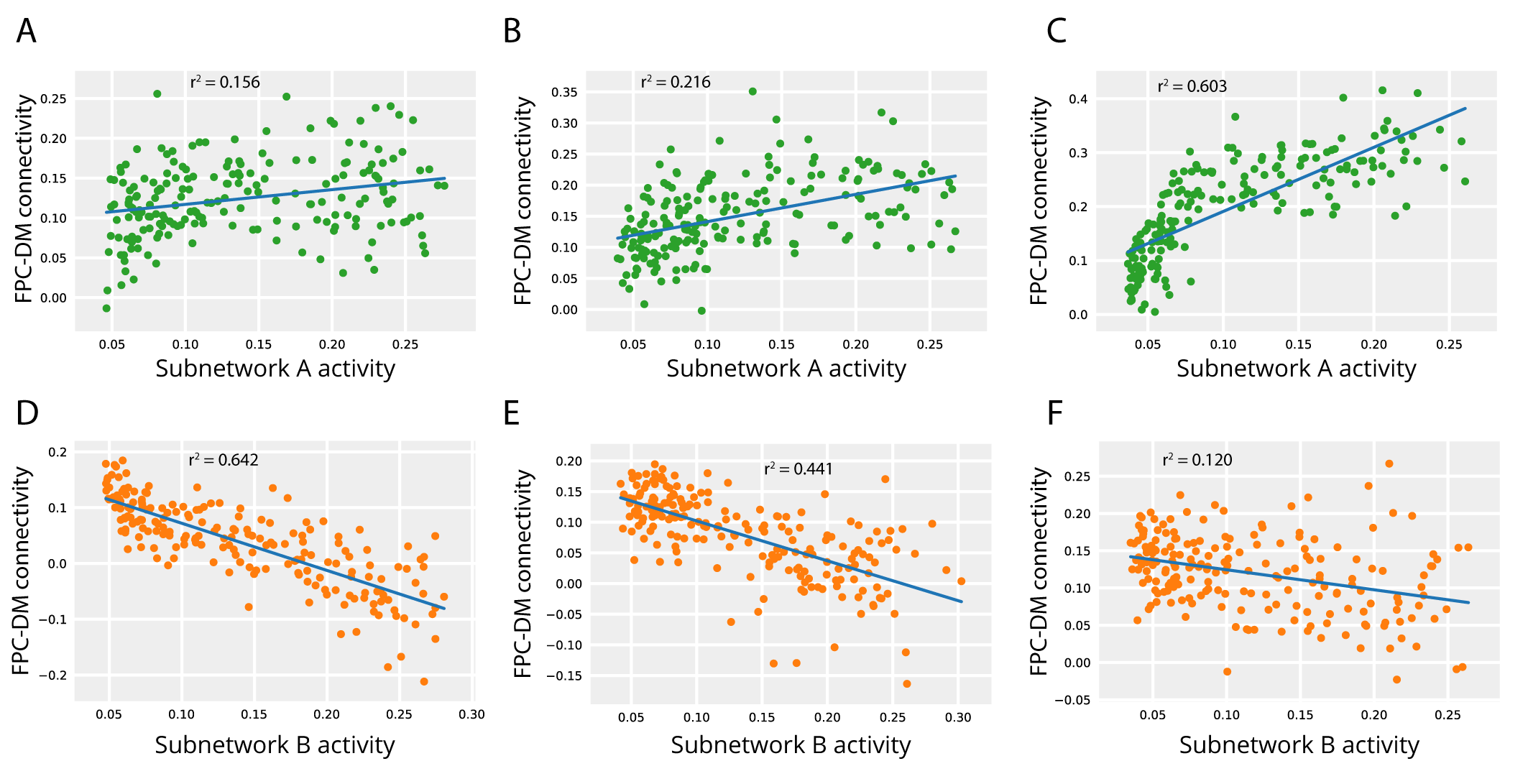}}
\caption{\textbf{Simulated frequency effect on activity-connectivity variance.} More variance of the relationship between subnetwork (A) activity and the functional coupling between the frontoparietal and default mode system is explained as the frequency of subnetwork (A) is increased in the dynamical model. However, less variance of the relationship between subnetwork (B) activity and the functional coupling between the frontoparietal and default mode systems is explained as the frequency of subnetwork (B) is increased in the dynamical model.}\label{figS14}
\end{figure}

\bibliographystyle{plain}